\documentclass[letter,aj]{emulateapj-rtx4}

\def\figwidth{6.7cm}

\def\medfigwidth{7.5cm}
\def\fullfigwidth{8.4cm}
\def\fullfigheight{11.76cm}

\usepackage{amssymb}
\usepackage{natbib}
\usepackage{amsmath}
\usepackage{graphicx}
\usepackage{longtable}

\begin{document}

\title{The Ages of Type Ia Supernova Progenitors}

\shorttitle{Progenitors of SNe Ia}
\shortauthors{Brandt et al.}

\author{Timothy D. Brandt\altaffilmark{1}, Rita Tojeiro\altaffilmark{2},
  \'Eric Aubourg\altaffilmark{3,4,1}, Alan Heavens\altaffilmark{5},
  Raul Jimenez\altaffilmark{6}, and Michael A. Strauss\altaffilmark{1}}

\altaffiltext{1}{Department of Astrophysical Sciences, Peyton Hall, Princeton University, Princeton, NJ 08544, USA}
\altaffiltext{2}{Institute of Cosmology and Gravitation, University of Portsmouth, Dennis Sciama Building, Burnaby Road, Portsmouth, PO1 3FX}
\altaffiltext{3}{Astroparticule et Cosmologie APC, UMR 7164,
  Universit\'e Paris Diderot, 10 rue Alice Domon et L\'eonie Duquet,
  75205 Paris cedex 13, France}
\altaffiltext{4}{CEA, Irfu, SPP, Centre de Saclay, 91191 Gif sur Yvette cedex, France}
\altaffiltext{5}{SUPA, Institute for Astronomy, University of Edinburgh, Royal Observatory, Blackford Hill, Edinburgh EH9-3HJ, UK}
\altaffiltext{6}{ICREA \& Institute for Sciences of the Cosmos (ICCUB), University of  
Barcelona, Barcelona 08028, Spain}

\begin{abstract}
Using light curves and host galaxy spectra of 101 Type Ia supernovae
(SNe Ia) with redshift $z \lesssim 0.3$ from the SDSS Supernova Survey
(SDSS-SN), we derive the SN Ia rate as a function of progenitor age
(the delay time distribution, or DTD).  We use the VESPA stellar
population synthesis algorithm to analyze the SDSS spectra of all
galaxies in the field searched by SDSS-SN, giving us a reference
sample of 77,000 galaxies for our SN Ia hosts.  Our method does not
assume any a priori shape for the DTD and therefore is minimally
parametric.  We present the DTD in physical units for high stretch
(luminous, slow declining) and low stretch (subluminous, fast
declining) supernovae in three progenitor age bins. We find strong
evidence of two progenitor channels: one that produces high stretch
SNe Ia $\lesssim 400$ Myr after the birth of the progenitor system,
and one that produces low stretch SNe Ia with a delay $\gtrsim 2.4$
Gyr.  We find that each channel contributes roughly half of the Type
Ia rate in our reference sample.  We also construct the average
spectra of high stretch and low stretch SN Ia host galaxies, and find
that the difference of these spectra looks like a main sequence B star
with nebular emission lines indicative of star formation.  This
supports our finding that there are two populations of SNe Ia, and
indicates that the progenitors of high stretch SNe are at the least
associated with very recent star formation in the last few tens of
Myr.  Our results provide valuable constraints for models of Type Ia
progenitors and may help improve the calibration of SNe Ia as standard
candles.
\end{abstract}

\keywords{Cosmology: Distance Scale, Galaxies: Stellar Content, Stars: Supernovae: General}


\section{Introduction}

Type Ia supernovae (SNe Ia) are thought to be the products of a
thermonuclear explosion of a white dwarf reaching the Chandrasekhar
mass - the mass limit allowed by the supporting electron degeneracy
pressure.  This requires the white dwarf to accrete mass from a binary
companion, though the nature of this companion and the timescale of
the accretion are hotly debated.  The {\it single degenerate} model
proposes the companion to be a main sequence or giant star with mass
transfer by Roche lobe overflow, while the {\it double degenerate}
scenario proposes that the companion is also a white dwarf and the two
objects merge.  While the connection between these two scenarios and
the delay time from progenitor birth to SN Ia explosion is unclear,
knowledge of this delay would constrain the initial masses of SN Ia
progenitors.  

Interest in SNe Ia has greatly increased since their use as
standardized candles led the discovery of the accelerated expansion of
the Universe
\citep{RiessEtAl98, PerlmutterEtAl99}.  The latest
generation of SN Ia surveys, including SNLS \citep{AstierEtAl06},
ESSENCE \citep{Wood-VaseyEtAl07}, and
SDSS-SN \citep{KesslerEtAl09}, however, are limited by the
possibility of
systematic uncertainties in their calibration of SN Ia luminosities.
This has sparked great interest in any physical properties that could
alter the peak luminosity/light curve width relation
\citep{Phillips93} used to calibrate SNe Ia. In particular, it has
highlighted our incomplete knowledge of SN Ia progenitor systems.  

Due to their low luminosities, SN Ia progenitors have never been
directly observed.  Constraints on their nature and rates of explosion
must therefore come from studies of SN Ia environments, and a variety
of techniques have been used to carry out such studies.
\cite{SullivanEtAl06}, \cite{MannucciEtAl06}, and others use host galaxy
photometry and \cite{Gallagheretal05} use spectra to derive stellar
populations and metallicities, and correlate these luminosity-weighted
average values against SN Ia rates and properties.
\citeauthor{SullivanEtAl06} and \citeauthor{MannucciEtAl06} see higher
SN Ia rates in star-forming galaxies, while
\citeauthor{Gallagheretal05} find that spirals host more luminous
SNe.  To derive constraints on the ages of the progenitor
systems, \cite{NeillEtAl06} and \cite{Gal-YamMaoz04} compare
the evolution of the supernova rate with the evolution of the cosmic
star formation rate.  Because this rate changes slowly with redshift
in the local universe, the method places only weak constraints on the
progenitor ages.  \cite{CooperEtAl09} have used host galaxy
clustering as a proxy for the metallicities of SN Ia environments and
found a significant correlation with supernova rate or luminosity.
\cite{BadenesEtAl09} have  measured the {\it local} stellar
populations for four historical SNe Ia in the Large Magellanic Cloud,
finding that three of the supernovae live in regions with little star
formation over the last several Gyr.  

Over the past decade and a half, analyses of SN Ia environments
(including several papers cited above) have produced evidence that
there are at least two distinct populations of SNe Ia:  blue,
star-forming galaxies have higher supernova rates and host more
luminous, slower-declining 
supernovae than do red, passive galaxies \citep[][but see
  \citealt{Schawinski09}]{HamuyEtAl96, Howell01,
  VanDenBerghEtAl05, MannucciEtAl05}.  More recently, this conclusion
has been confirmed in large, local SN Ia samples by
\cite{HickenEtAl09a} and \cite{MaozEtAl10} and in a higher
redshift sample by \cite{SullivanEtAl10}.  The observation of at
least two populations has led to the so-called
$A+B$ model \citep{ScannapiecoBildsten05}, in which the supernova
rate is modeled as the sum of a term proportional to total
stellar mass (the delayed component) and a term proportional to recent
star formation (the prompt component):
\begin{equation} 
\mathit{SNR} = AM_* + B\dot{M}_*.
\label{eq:AB_model}
\end{equation}
Values of $A$ and $B$ have been determined by \cite{SullivanEtAl06},
\cite{NeillEtAl06}, \cite{MannucciEtAl05} and others, with significant
scatter between the 
groups due to different SN Ia samples, methods for deriving $A$ and
$B$, definitions of stellar masses and proxies for the star formation
rates.  \cite{AubourgEtAl08} used host galaxy spectra and found
that the supernova rate per unit mass for young stars is $\sim 500$ times
higher than for old stellar populations, a factor of $\sim 5$ higher than
earlier results.  They also found evidence that the time scale of the
prompt component is $\lesssim 180$ Myr.  

Unfortunately, the $A+B$ model provides only weak constraints on the
ages associated with the prompt and delayed components.  Because of
the strong dependence of stellar evolution timescales on initial mass,
more precise ages can powerfully constrain the main-sequence masses
of the progenitors.  This will require a better approximation to the
full delay-time distribution (DTD), the explosion rate as a function
of progenitor age.  The DTD, $\epsilon(t)$, has units of number of
supernovae per unit stellar mass per year, and relates a galaxy's
supernova rate to its star formation rate $\psi(t)$:
\begin{equation}
\mathit{SNR}(t) = \int_0^{t} \epsilon(t-t') \psi(t')\,\mathrm{d}t'.
\end{equation}
Thus, $\epsilon(t)$ represents
the probability per year that a unit of stellar mass will produce a
Type Ia supernova a time $t$ after its formation.  The DTD sets the
rates and timescales of SN Ia production that must be matched by
progenitor models.  

There have been several attempts to parametrize and measure the
DTD. \cite{PritchetEtAl08} assume a continuous, power-law DTD
($\epsilon(t) \propto t^{-p}$) and find that the SNLS survey constrains
$p$ to lie in the range $0.3 \leq p \leq 0.7$.  Roughly speaking, this
is required for consistency with measurements of $A$ and $B$ from,
e.g., \cite{SullivanEtAl06}.  Using
a theoretical argument to calculate the rate of white dwarf formation
per unit stellar mass, they find that a single-degenerate model can
yield a power law DTD of this form only if the fraction of white
dwarfs which explode as SNe Ia is
independent of progenitor mass.  Because less massive progenitors
should produce less massive remnants, \citeauthor{PritchetEtAl08}
conclude that there must be another, possibly double degenerate,
route to SNe Ia. \cite{TotaniEtAl08} constrain the DTD in old stellar
populations by selecting SN candidates in passive galaxies from the
Subaru/XMM-Newton Deep Survey \citep[SXDS,][]{FurusawaEtAl08}.  Combining these high redshift
observations with the local observed SN Ia rate in
ellipticals, they find $\epsilon(t) \propto t^{-1}$ in the
range $0.1 < t < 10$ Gyr.  \citeauthor{TotaniEtAl08} argue that a
DTD of this form supports a double-degenerate origin for delayed SNe
Ia.  

In this paper, we aim to provide better constraints on the DTD and to
determine whether luminous SNe Ia (with temporally ``stretched''
light curves) and less luminous events have different progenitors from
one another.  
Any attempt to measure the DTD relies on the intrinsic range of galaxy
properties, and in particular, the differences between supernova hosts
and non-hosts.  It is therefore essential to have a supernova sample
with a well-understood selection function and a well-defined control
group of galaxies monitored for SNe.  Otherwise, differences between
hosts and non-hosts may be attributed to biases or survey systematics
rather than to supernova rates.  Nearby surveys often comprise
SNe discovered by many different techniques, making it difficult to
define or build a control sample.  High redshift searches are
often blind (i.e.~sensitive to SNe in all galaxies in the field), but
lack spectra of the field galaxies.  

The Sloan Digital Sky Survey Supernova Survey \citep[SDSS-SN,
][]{FriemanEtAl08, SakoEtAl08} satisfies both
criteria.  It is the only controlled, blind, difference 
imaging supernova search at low redshift, with most supernovae in the
range $0.05 \lesssim z \lesssim 0.35$.  SDSS
\citep{YorkEtAl00} also has a large spectroscopic
catalog of galaxies, of which about 83,000 lie in the stripe of sky
monitored for supernovae.  SDSS-SN is therefore the ideal survey with
which to perform a statistical comparison of the spectroscopic
properties of hosts and non-hosts.  

For our study we select 101 SNe Ia from the SDSS-SN sample with
acceptable light curves and SDSS host galaxy spectra.  We then use VErsatile
SPectral Analysis \citep{TojeiroEtAl07} - VESPA - to derive star
formation histories for all galaxies in the control sample.  Given
any assumed delay time distribution, these star formation histories
allow us to simulate a population of hosts.  We then use the spectra
themselves to compare these mock hosts to the observed SN Ia host
galaxies.  We compute likelihoods for a large number of assumed delay
time distributions, thereby constraining the DTD.  

We have organized this paper as follows: in Section \ref{sec:data} we
describe our datasets; in Sections \ref{sec:method} and
\ref{sec:constrain} we describe our
methodology; in Section \ref{sec:results} we present our results and
we finally discuss and conclude in Section \ref{sec:discussion}. 

\section{Data: SDSS-SN} \label{sec:data}

The SDSS Supernova Survey (SDSS-SN) was carried out along ``Stripe
82'' \citep{YorkEtAl00}, a
region of sky 2.5 degrees wide centered on the Celestial Equator,
stretching from $20.7$ hours to $3.9$ hours ($-50^\circ$ - $+59^\circ$)
Right Ascension for a total area of 270 deg$^2$.  It was observed up
to 80 times with the SDSS imaging camera \citep{GunnEtAl98};
supernova candidates were identified by difference imaging, and
followed up with spectroscopy on other telescopes, and photometry from
the SDSS telescope \citep{GunnEtAl06} and other telescopes as
well.  

SDSS-SN is described in detail in \cite{FriemanEtAl08} and
\cite{SakoEtAl08}.  Operating for three months of the year
from 2005 to 2007, SDSS-SN has identified
over 400 spectroscopically confirmed SNe Ia in the redshift range $0.05
\lesssim z \lesssim 0.35$, of which 146 from the 2005 season have
$ugriz$ light curves published in \cite{HoltzmanEtAl08}.
Preliminary data for the remainder were kindly provided to us by the
SDSS supernova team.  We have selected definitive and probable SNe Ia
this catalog of preliminary photometry using the types reported
through the Central Bureau for Astronomical
Telegrams\footnote{http://www.cfa.harvard.edu/iau/lists/Supernovae.html}.
We have further excluded two objects listed as peculiar SNe Ia, 
2005hk and 2007qd. 

The preliminary photometric data lack errorbars, which we have
estimated using the observed correlation between errors and magnitudes
in each band in the published 2005 data \citep{HoltzmanEtAl08}.
Because our final analysis only uses the light curves to divide our
sample into two broad subsamples, our results are insensitive to the
details of these error estimates.  We have also checked the 
reliability of the preliminary flux measurements using objects from
the 2005 season for which final data are also available. 
The differences between preliminary and final published fluxes are
approximately Cauchy distributed, with a mean offset from zero of
$\sim 0.5~\mu{}\mathrm{Jy}$ and a full width at half maximum of $\sim
2.5~\mu{}\mathrm{Jy}$.  These are much fainter than a typical SDSS
supernova peak magnitude of $\lesssim 20$, or a peak flux of $\gtrsim
40~\mu{}\mathrm{Jy}$.  


The SDSS carried out spectroscopy of galaxies on Stripe 82, using the
standard selection algorithms used throughout the SDSS survey.  These
include the Main Galaxy selection described in
\cite{StraussEtAl02}, consisting of all galaxies with
\cite{Petrosian76} $r$-band magnitudes brighter than
17.77 and with a median redshift of order 0.13, and the Luminous Red
Galaxy (LRG) sample of \cite{EisensteinEtAl01}, which selects the most
luminous red ellipticals to $z\sim 0.55$.  Because of the limited
imaging carried out by SDSS in the Fall months, however, additional
spectroscopy was carried out on Stripe 82 using a variety of
algorithms that went beyond the main spectroscopic sample
\citep{Adelman-McCarthyEtAl06}, including  a sample designed to calibrate
photometric redshifts, a sample of low-luminosity galaxies, an
extension of the LRG sample to fainter and bluer objects, and a sample
flux-limited in the $u$-band.  In all, there are about 83,000 unique
galaxies on Stripe 82 with spectroscopic data, of which only 23,000
were selected using the main survey algorithms.  We have further
excluded all galaxies with bad redshifts and, because we are
interested in differences only in the {\it stellar} populations
between SN Ia hosts and non-hosts, galaxies flagged as AGNs or QSOs by
the SDSS pipeline.  This leaves about 77,000 unique galaxies from the
Seventh Data Release of SDSS \citep{AbazajianEtAl09}, of which 22,000
are in the Main Galaxy Sample. 

We have matched the SDSS-SN sample with this spectroscopic sample,
yielding a total of 133 SNe with host spectra in SDSS.  The
light curves of these supernovae range in quality, but 101 are
sufficiently good to measure parameters such as the peak flux and decline
rate (see \S\ref{sec:stretch} and Figure \ref{fig:lc_samples} for
examples).  The 77,000 unique galaxies in Stripe 82 were not uniformly
selected, and do not represent a flux-limited sample.  However, our
subsample of 101 SN hosts was selected from these 77,000 galaxies
based solely on the occurrence of a detectable SN Ia.  These 101 SNe
with SDSS hosts form the sample used in the rest of our analysis.  

\section{Methodology: Initial Data Analysis} \label{sec:method}

To measure the delay time distribution, we need to be able to derive
star formation histories for galaxies from their SDSS spectra.
Because of the strong association of luminous (high stretch) SNe Ia
with recent star formation seen by \cite{SullivanEtAl06} and others,
we also investigate the dependence of the delay time distribution on
supernova stretch.  We begin by describing how we derive the star
formation history and stretch from host galaxy spectrum and SN light
curve, respectively.  We then discuss
the division of SNe Ia into a high stretch and a low stretch subgroup,
and finally introduce the average spectra of Type Ia supernova hosts.  

\subsection{A star formation history in three age bins} \label{sec:vespa}

We use VESPA \citep{TojeiroEtAl07}
to model the observed spectrum of a galaxy as a
linear combination of up to sixteen single stellar populations (SSPs) of
different ages and metallicities shielded by a common dust screen.
The resulting ages, stellar masses formed, metallicities, and dust
values for roughly 800,000 individual galaxies in SDSS's Seventh Data
Release \citep{AbazajianEtAl09} have been published in
\cite{TojeiroEtAl09}. This public
database currently holds only the SDSS Main Galaxy and LRG samples; the
additional Stripe 82 galaxies used in this paper will be added
soon. 

While VESPA recovers the star formation history in up to sixteen
logarithmically spaced age bins, the number and temporal width of
these bins vary according to the quality of a given galaxy's
spectrum.  VESPA's analysis and solutions are also
model-dependent, particularly for recent star formation
\citep{TojeiroEtAl09}.  The recovered mass of stars formed in the most
recent age bins, more recently than 70 Myr ago (in the
rest-frame of the galaxy) with our choice of bins, is particularly
sensitive to the dust modeling.  The mass formed in the next group of
age bins, between 70 and 420 Myr, is sensitive to the choice of
SSP \citep[see Figures 15 and 20 of][]{TojeiroEtAl09}. To deal with
these limitations, we:
\begin{itemize}
\item select the stellar models that give the most physically
  reasonable answer, and
\item limit our recovered star formation histories to three age bins.
\end{itemize}
Over a typical ensemble of galaxies, the recovered star formation rate
from 70 to 420 Myr is anticorrelated with and significantly lower than
the rate up to 70 Myr.  The difficulty of determining the abundance of
a few hundred Myr old stars in the presence of very recent star
formation is less pronounced with the \cite{Maraston05} SSP models.
To limit the impact of this, we combine all age bins younger than 420
Myr into a single bin.  With these choices, the recovered star
formation rate becomes a smooth function of lookback time.  We seek
the star formation histories of all galaxies in the same bins, and
therefore degrade the age bins older than 420 Myr into two broad bins.
We thus recover the star formation histories for all galaxies in the
three age bins detailed in Table \ref{tab:bin_info}.  For each galaxy,
we obtain the stellar mass formed over each age range, the
mass-weighted metallicity in each bin, and an average value for the
optical depth of interstellar dust.  

The boundaries between the age bins in Table \ref{tab:bin_info} are
chosen for convenience and to correspond to the main sequence
lifetimes of stellar spectral classes: we take them as given and
estimate stellar masses.  This does not imply that the exact age
boundaries have any physical significance; they are simply
logarithmically spaced divisions.  

\begin{deluxetable}{ccc}
\tablewidth{0pt}
\tablecaption{TABLE \ref{tab:bin_info} \\ The three VESPA age bins.  }
\tablehead{
\colhead{Bin} & \colhead{Age Range (Gyr)} & \colhead{MS Spectral
  Types\tablenotemark{a}}
}
\startdata
1 & 0.002 -- 0.42  & O and B \\
2 & 0.42 -- 2.4 & A \\
3 & 2.4 -- 14 & F and later 
\enddata
\tablenotetext{a}{The main
  sequence (MS) spectral types that dominate the galaxy spectra in
  this age range.}
\label{tab:bin_info}
\end{deluxetable}%

\subsection{Stretch measurements} \label{sec:stretch}

We have refitted all the light curves from SDSS-SN
(\S\ref{sec:data}) with the SALT II algorithm \citep{GuyEtAl07}.
This template-based routine takes the redshift and all measured fluxes
in several bands as input, and performs a four-parameter fit.  We fit
to the SDSS-SN light curves in the $g$, $r$, and $i$ bands.  We
exclude the $u$ band data from all analyses because of its poor
signal-to-noise ratio, and exclude the $z$ band data because 
Type Ia near-infrared light curves are often double-peaked and poorly
fit by the SALT II templates.  The SALT II outputs consist of
a stretch-like parameter ($x_1$ in \citealt{GuyEtAl07}), color $c$
($B-V$ at $B$ band maximum relative to \citeauthor{GuyEtAl07}'s sample average),
rest-frame peak B magnitude $m_B$, and time of maximum light.  For this
work, we convert $x_1$ back to the traditional stretch $s$ using the
relation given in \cite{GuyEtAl07}:  
\begin{equation}
s = 0.98 + 0.091x_1 + 0.003x_1^2 - 0.00075x_1^3.
\label{eq:x1_s_def}
\end{equation}
Stretch $s$ is a dimensionless parameter that describes the temporal
width of the Type Ia light curve relative to a fiducial average.
In our sample, $s$ ranges from $0.70$ to $1.31$ with a median of
$0.94$.  The color parameter $c$ is a degenerate combination of
host galaxy extinction and intrinsic color, which for our SNe ranges
from $-0.14$ to $0.90$ with a median of $0.06$. 

\begin{figure}
\begin{center}
\includegraphics[width=\fullfigwidth]{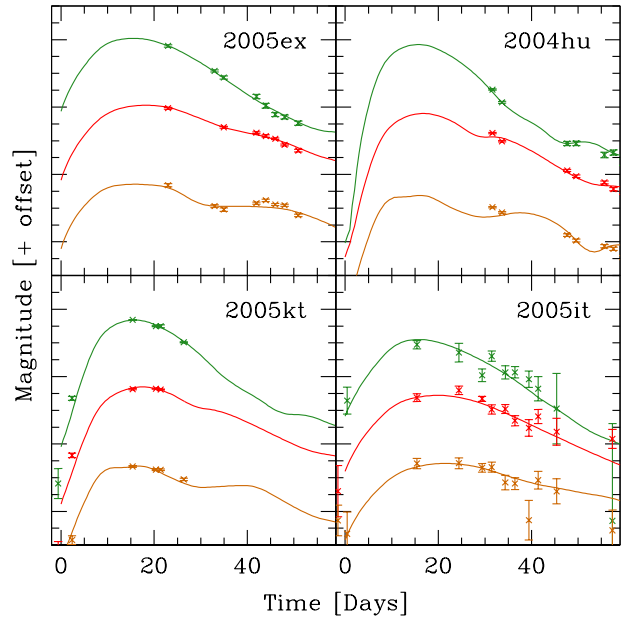}
\end{center}
\caption{Light curves in $g$, $r$, and $i$ (top to bottom) for four
  objects with host galaxies in the SDSS spectroscopic
  sample; the bands are offset for clarity.  The object shown in the
  upper right panel has little data near peak light and was therefore
  excluded as having an uncertain stretch.  The other three objects
  entered our final sample of 101 SNe.  Even though the object shown
  in the top left panel, 2005ex, has no data before maximum light, it
  is very well-sampled and well-fit.  Because we only use these fits
  to divide our sample in two (\S\ref{sec:stretchdiv}), we want to
  maximize our sample size by including SNe like 2005ex.}
\label{fig:lc_samples}
\end{figure}

While SALT II estimates the errors of all of its fitted
parameters, we have assigned subjective quality ratings to each
object as an independent error metric, based on our confidence in the
best-fit value of the stretch parameter $s$.  We have excluded
objects with poorly determined stretches (32 of the 133 with host-galaxy
spectra) from all further analysis.  Four sample light curves in $g$,
$r$, and $i$, including one that was rejected in this fashion, are
shown in Figure \ref{fig:lc_samples}.  We thus
arrive at our sample of 101 SNe selected from about 77,000 control
galaxies with spectra.  

The selection of these 101 SNe Ia by inspection of the light curves
could possibly introduce a bias in our results.  To minimize the
likelihood of this, we made all assessments of light curve quality
blindly, with no knowledge of the host galaxy properties.  Only after
all of our supernova light curves were fitted and rated did we perform
any analysis on their hosts.  We stress that we do not calibrate SNe
as standard candles (which would require the recovery of three
parameters from the light curve fit), but only use the light curves to
divide our sample in two (\S\ref{sec:stretchdiv}).  Because of this,
our light curves need not be as well-sampled as for cosmological
analyses. We present details of this final sample of 101 SNe in Table
\ref{tab:sne}. 

\begin{deluxetable*}{llllllll}
\tablewidth{0pt}
\tablenum{2}
\tablecolumns{7}
\tablecaption{TABLE \ref{tab:sne} \\ The Restricted SDSS-SN Sample}
\tablehead{\colhead{IAU ID}                                      & 
           \colhead{Host}                                        & 
	   \multicolumn{2}{c}{RA (J2000) Dec\,\tablenotemark{a}} & 
	   \colhead{Stretch}                                     & 
	   \multicolumn{3}{c}{VESPA Host Stellar Mass (M$_\odot$)} \\
	   \colhead{}                                            &
	   \colhead{Redshift}                                    &
	   \colhead{{h}\phn{m}\phn{s}}                           &
	   \colhead{\phn{\arcdeg}~\phn{\arcmin}~\phn{\arcsec}}   &
	   \colhead{}                                            &
	   \colhead{2 - 420 Myr}                                 &
	   \colhead{0.42 - 2.4 Gyr}                              &
	   \colhead{2.4 - 14 Gyr}                                }
\startdata
2004hz &  0.1427 & 02 05 03.0 & +0 50 11.9 & high & $8.8 \times 10^{8}$ & 0 & $1.6 \times 10^{11}$  \\ 
2004ia &  0.1437 & 02 18 02.3 & +0 33 32.4 & high & $7.2 \times 10^{10}$ & 0 & $2.6 \times 10^{11}$  \\ 
2004ie &  0.0513 & 22 01 46.6 & +1 14 12.0 & high & $3.7 \times 10^{7}$ & $9.9 \times 10^{8}$ & $2.9 \times 10^{10}$  \\ 
2004ig &  0.1830 & 00 05 51.8 & +0 59 45.1 & high & $3.2 \times 10^{9}$ & $2.9 \times 10^{10}$ & $1.4 \times 10^{11}$  \\ 
2005ed &  0.0865 & 00 02 49.4 & +0 45 04.3 & low  & 0 & $8.8 \times 10^{9}$ & $4.0 \times 10^{11}$  \\ 
2005ef &  0.1077 & 00 58 22.9 & +0 40 44.4 & low  & $3.5 \times 10^{8}$ & $9.5 \times 10^{9}$ & $3.4 \times 10^{11}$  \\ 
2005eg &  0.1914 & 01 02 08.5 & +0 52 44.4 & high & $1.5 \times 10^{9}$ & $4.6 \times 10^{10}$ & $3.8 \times 10^{11}$  \\ 
2005ex &  0.0935 & 01 41 51.2 & +0 52 35.0 & high & $1.8 \times 10^{9}$ & 0 & $1.7 \times 10^{11}$  \\ 
2005ez &  0.1298 & 03 07 11.0 & +1 07 10.4 & low  & $3.4 \times 10^{9}$ & 0 & $1.5 \times 10^{11}$  \\ 
2005fa &  0.1615 & 01 39 36.1 & +0 45 31.5 & high & 0 & $1.9 \times 10^{10}$ & $6.3 \times 10^{11}$  \\ 
2005fh &  0.1190 & 23 17 29.7 & +0 25 45.8 & high & 0 & $4.9 \times 10^{10}$ & $2.0 \times 10^{11}$  \\ 
2005fv &  0.1181 & 03 05 22.4 & +0 51 30.1 & high & $4.4 \times 10^{8}$ & $4.7 \times 10^{10}$ & $2.5 \times 10^{11}$  \\ 
2005gb &  0.0864 & 01 16 12.6 & +0 47 31.0 & low  & $4.3 \times 10^{8}$ & $1.3 \times 10^{10}$ & $3.7 \times 10^{11}$  \\ 
2005gh &  0.2591 & 20 50 36.4 & +0 21 14.8 & high & $1.3 \times 10^{9}$ & $7.4 \times 10^{10}$ & $7.5 \times 10^{11}$  \\ 
2005gp &  0.1266 & 03 41 59.3 & +0 46 57.6 & low  & $7.4 \times 10^{8}$ & $4.6 \times 10^{9}$ & $7.3 \times 10^{10}$  \\ 
2005hc &  0.0459 & 01 56 47.9 & +0 12 49.2 & high & 0 & $7.4 \times 10^{9}$ & $3.9 \times 10^{11}$  \\ 
2005hj &  0.0574 & 01 26 48.3 & $-$1 14 16.8 & high & $9.8 \times 10^{7}$ & $6.0 \times 10^{8}$ & $1.9 \times 10^{10}$  \\ 
2005hn &  0.1085 & 21 57 04.2 & +0 13 24.5 & low  & $6.2 \times 10^{8}$ & 0 & $1.1 \times 10^{10}$  \\ 
2005ho &  0.0628 & 00 59 24.1 & +0 00 09.4 & high & $2.3 \times 10^{9}$ & 0 & $2.3 \times 10^{10}$  \\ 
2005if &  0.0670 & 03 30 12.9 & +0 58 28.5 & low  & $3.1 \times 10^{9}$ & 0 & $1.1 \times 10^{11}$  \\ 
2005ij &  0.1245 & 03 04 21.3 & $-$1 03 46.6 & high & $2.7 \times 10^{9}$ & $2.0 \times 10^{10}$ & $1.2 \times 10^{11}$  \\ 
2005ir &  0.0763 & 01 16 43.8 & +0 47 40.4 & high & $3.1 \times 10^{9}$ & $9.3 \times 10^{9}$ & $2.9 \times 10^{11}$  \\ 
2005je &  0.0939 & 02 35 26.6 & +1 04 29.6 & low  & 0 & $1.9 \times 10^{10}$ & $8.1 \times 10^{11}$  \\ 
2005js &  0.0796 & 01 34 41.5 & +0 36 19.3 & low  & 0 & $1.1 \times 10^{10}$ & $5.7 \times 10^{11}$  \\ 
2005kt &  0.0653 & 01 10 58.0 & +0 16 34.1 & low  & 0 & $1.9 \times 10^{10}$ & $1.3 \times 10^{11}$  \\ 
2005ku &  0.0454 & 22 59 42.6 & +0 00 49.3 & high & $3.0 \times 10^{9}$ & 0 & $5.3 \times 10^{10}$  \\ 
2005lj &  0.0777 & 01 57 43.0 & +0 10 46.0 & high & $5.9 \times 10^{8}$ & 0 & $5.2 \times 10^{10}$  \\ 
2005lk &  0.1042 & 21 59 49.4 & $-$1 11 37.3 & low  & $7.1 \times 10^{8}$ & $4.0 \times 10^{10}$ & $7.9 \times 10^{11}$  \\ 
2006eq &  0.0494 & 21 28 37.1 & +1 13 41.2 & low  & $9.4 \times 10^{7}$ & $8.3 \times 10^{9}$ & $2.1 \times 10^{11}$  \\ 
2006er &  0.0843 & 00 21 37.5 & $-$1 00 35.9 & low  & 0 & $1.0 \times 10^{10}$ & $5.0 \times 10^{11}$  \\ 
2006ex &  0.1472 & 20 38 43.9 & +0 28 28.3 & high & $1.3 \times 10^{9}$ & $1.1 \times 10^{11}$ & $2.7 \times 10^{11}$  \\ 
2006fb &  0.2451 & 23 35 51.5 & +0 10 37.6 & high & $1.0 \times 10^{9}$ & $4.2 \times 10^{10}$ & $3.6 \times 10^{11}$  \\ 
2006fd &  0.0799 & 20 37 53.2 & +1 13 16.1 & low  & 0 & $2.2 \times 10^{10}$ & $2.6 \times 10^{11}$  \\ 
2006ff &  0.2353 & 00 26 35.6 & +0 18 07.5 & high & $2.4 \times 10^{9}$ & $6.6 \times 10^{9}$ & $3.2 \times 10^{11}$  \\ 
2006fi &  0.2306 & 22 19 50.3 & +0 01 27.8 & high & $4.1 \times 10^{9}$ & $5.8 \times 10^{10}$ & $2.3 \times 10^{11}$  \\ 
2006fl &  0.1717 & 22 11 27.7 & +0 45 21.5 & high & $8.7 \times 10^{8}$ & $9.1 \times 10^{9}$ & $6.2 \times 10^{10}$  \\ 
2006fs &  0.0991 & 21 09 59.0 & +0 24 31.6 & high & $2.3 \times 10^{9}$ & $2.1 \times 10^{10}$ & $4.1 \times 10^{11}$  \\ 
2006fu &  0.1985 & 23 51 08.4 & +0 44 46.9 & high & $1.5 \times 10^{9}$ & $2.5 \times 10^{8}$ & $4.0 \times 10^{10}$  \\ 
2006fv &  0.1319 & 01 21 37.9 & +0 24 52.2 & low  & 0 & $8.9 \times 10^{9}$ & $5.0 \times 10^{11}$  \\ 
2006fy &  0.0827 & 23 26 40.2 & +0 50 24.9 & high & $2.5 \times 10^{9}$ & $1.2 \times 10^{9}$ & $3.5 \times 10^{10}$  \\ 
2006fz &  0.1047 & 00 16 41.4 & +0 25 28.3 & low  & 0 & $2.1 \times 10^{10}$ & $9.6 \times 10^{11}$  \\ 
2006gb &  0.2660 & 23 59 16.5 & $-$1 15 01.3 & low  & $4.5 \times 10^{9}$ & $6.5 \times 10^{10}$ & 0  \\ 
2006gx &  0.1807 & 02 48 14.1 & +0 20 49.3 & high & $6.5 \times 10^{9}$ & $8.7 \times 10^{9}$ & $5.3 \times 10^{10}$  \\ 
2006hd &  0.2983 & 21 44 03.5 & +0 43 34.6 & high & $2.0 \times 10^{9}$ & $2.9 \times 10^{10}$ & $6.7 \times 10^{10}$  \\ 
2006hh &  0.2374 & 02 42 27.0 & +0 47 38.9 & low  & $2.9 \times 10^{8}$ & $6.9 \times 10^{10}$ & $1.2 \times 10^{12}$  \\ 
2006hr &  0.1576 & 01 50 15.6 & +0 53 14.1 & high & $1.4 \times 10^{10}$ & 0 & $8.4 \times 10^{10}$  \\ 
2006hw &  0.1394 & 03 13 03.4 & +0 28 17.9 & high & $2.4 \times 10^{10}$ & 0 & $3.9 \times 10^{11}$  \\ 
2006hx &  0.0454 & 01 13 57.3 & +0 22 18.0 & high & $1.3 \times 10^{8}$ & $1.6 \times 10^{10}$ & $1.6 \times 10^{11}$  \\ 
2006ia &  0.1749 & 02 07 19.2 & +1 15 07.5 & low  & $1.5 \times 10^{8}$ & $4.5 \times 10^{10}$ & $1.9 \times 10^{12}$  \\ 
2006ib &  0.1811 & 03 16 11.8 & +0 36 03.4 & low  & 0 & 0 & $1.4 \times 10^{11}$  \\ 
2006ju &  0.1486 & 23 24 39.0 & +0 43 06.0 & low  & $3.5 \times 10^{9}$ & $4.1 \times 10^{10}$ & $8.4 \times 10^{11}$  \\ 
2006jw &  0.2495 & 02 23 22.3 & +0 49 08.4 & high & $9.7 \times 10^{8}$ & $8.1 \times 10^{10}$ & $2.4 \times 10^{11}$  \\ 
2006jz &  0.1994 & 00 11 24.8 & +0 42 09.8 & low  & 0 & $1.1 \times 10^{11}$ & $1.7 \times 10^{12}$  \\ 
2006kd &  0.1363 & 01 07 50.0 & +0 49 41.5 & high & $7.7 \times 10^{8}$ & 0 & $1.2 \times 10^{11}$  \\ 
2006kq &  0.1983 & 21 15 36.6 & +0 19 17.1 & high & $2.0 \times 10^{9}$ & $4.0 \times 10^{10}$ & $1.4 \times 10^{11}$  \\ 
2006kw &  0.1854 & 02 14 58.0 & +0 36 09.0 & high & $9.6 \times 10^{9}$ & 0 & $6.2 \times 10^{10}$  \\ 
2006kx &  0.1599 & 03 42 14.7 & +0 28 41.8 & high & $4.8 \times 10^{7}$ & $8.2 \times 10^{9}$ & $5.5 \times 10^{10}$  \\ 
2006lb &  0.1819 & 03 19 28.2 & +0 19 04.9 & high & $3.0 \times 10^{9}$ & $8.3 \times 10^{9}$ & $5.1 \times 10^{10}$  \\ 
2006nd &  0.1288 & 22 44 59.1 & $-$1 00 23.8 & high & $2.4 \times 10^{10}$ & 0 & $5.5 \times 10^{11}$  \\ 
2006ne &  0.0466 & 01 13 37.8 & +0 25 25.9 & low  & $1.6 \times 10^{9}$ & $1.1 \times 10^{10}$ & $1.7 \times 10^{11}$  \\ 
2006ni &  0.1750 & 20 54 52.4 & +0 11 41.4 & low  & 0 & $6.0 \times 10^{10}$ & $9.3 \times 10^{11}$  \\ 
2006nn &  0.1969 & 01 45 41.0 & $-$1 03 15.8 & high & $1.3 \times 10^{9}$ & 0 & $1.3 \times 10^{11}$  \\ 
2006nz &  0.0381 & 00 56 29.2 & $-$1 13 36.1 & low  & 0 & $1.6 \times 10^{9}$ & $7.6 \times 10^{10}$  \\ 
2006oa &  0.0625 & 21 23 42.9 & +0 50 36.5 & high & 0 & $1.7 \times 10^{7}$ & $8.8 \times 10^{7}$  \\ 
2006ob &  0.0592 & 01 51 48.1 & +0 15 48.3 & low  & 0 & $2.1 \times 10^{11}$ & $8.8 \times 10^{11}$  \\ 
2006ol &  0.1191 & 23 28 07.2 & +0 51 22.9 & low  & 0 & $3.3 \times
10^{10}$ & $1.1 \times 10^{12}$  \\ \\ \\\\\\\\\\\\\\\\\\\\\\\\
& & & & & & 
\enddata
\tablenotetext{a}{Position of SN; data available from
  http://www.cfa.harvard.edu/iau/lists/Supernovae.html and
  http://www.cfa.harvard.edu/iau/cbat.html.}
\label{tab:sne}
\end{deluxetable*}%

\begin{deluxetable*}{llllllll}
\tablewidth{0pt}
\tablenum{2}
\tablecolumns{7}
\tablecaption{TABLE \ref{tab:sne}--{\it Continued}}
\tablehead{\colhead{IAU ID}                                      & 
           \colhead{Host}                                        & 
	   \multicolumn{2}{c}{RA (J2000) Dec\,\tablenotemark{a}} & 
	   \colhead{Stretch}                                     & 
	   \multicolumn{3}{c}{VESPA Host Stellar Mass (M$_\odot$)} \\
	   \colhead{}                                            &
	   \colhead{Redshift}                                    &
	   \colhead{{h}\phn{m}\phn{s}}                           &
	   \colhead{\phn{\arcdeg}~\phn{\arcmin}~\phn{\arcsec}}   &
	   \colhead{}                                            &
	   \colhead{2 - 420 Myr}                                 &
	   \colhead{0.42 - 2.4 Gyr}                              &
	   \colhead{2.4 - 14 Gyr}                                }
\startdata
2006on &  0.0719 & 21 55 58.5 & $-$1 04 12.7 & high & 0 & $9.8 \times 10^{9}$ & $4.7 \times 10^{10}$  \\ 
2006op &  0.0341 & 21 21 31.9 & +0 59 35.9 & low  & $2.4 \times 10^{8}$ & $3.7 \times 10^{9}$ & $2.8 \times 10^{10}$  \\ 
2006pe &  0.1611 & 00 23 09.2 & +0 03 13.1 & low  & 0 & $3.1 \times 10^{10}$ & $1.1 \times 10^{12}$  \\ 
2006py &  0.0578 & 22 41 42.0 & +0 08 12.9 & low  & 0 & $3.1 \times 10^{9}$ & $9.5 \times 10^{10}$  \\ 
2007ht &  0.0727 & 00 34 33.8 & $-$1 13 03.1 & low  & $1.4 \times 10^{8}$ & $1.6 \times 10^{10}$ & $6.8 \times 10^{11}$  \\ 
2007hx &  0.0794 & 02 06 27.1 & +0 53 58.3 & high & $5.4 \times 10^{9}$ & $1.8 \times 10^{10}$ & $3.4 \times 10^{11}$  \\ 
2007hy &  0.1814 & 03 39 42.3 & +1 05 32.2 & high & 0 & $2.2 \times 10^{10}$ & $5.3 \times 10^{11}$  \\ 
2007hz &  0.1393 & 21 03 08.9 & $-$1 01 45.1 & high & $7.1 \times 10^{9}$ & 0 & $6.7 \times 10^{11}$  \\ 
2007ia &  0.1310 & 03 43 10.1 & +0 06 08.9 & low  & $1.6 \times 10^{9}$ & $6.0 \times 10^{10}$ & $5.1 \times 10^{11}$  \\ 
2007id &  0.1603 & 21 46 00.5 & $-$1 13 03.9 & high & 0 & $4.2 \times 10^{9}$ & $1.8 \times 10^{11}$  \\ 
2007ie &  0.0934 & 22 17 36.7 & +0 36 48.0 & low  & $7.9 \times 10^{8}$ & 0 & 0  \\ 
2007jk &  0.1829 & 02 55 05.6 & +0 08 50.8 & high & $3.7 \times 10^{8}$ & $6.3 \times 10^{9}$ & $1.3 \times 10^{11}$  \\ 
2007js &  0.1464 & 20 36 48.7 & +0 05 54.4 & high & $9.4 \times 10^{9}$ & 0 & 0  \\ 
2007jt &  0.1447 & 02 28 32.8 & $-$1 02 31.6 & low  & $1.7 \times 10^{9}$ & $1.8 \times 10^{10}$ & $9.0 \times 10^{10}$  \\ 
2007kl &  0.2571 & 02 44 50.9 & +0 21 53.4 & high & $9.6 \times 10^{8}$ & $5.7 \times 10^{10}$ & $8.5 \times 10^{11}$  \\ 
2007kv &  0.3295 & 01 10 15.8 & +0 28 19.3 & high & $3.6 \times 10^{8}$ & $3.3 \times 10^{10}$ & $3.0 \times 10^{11}$  \\ 
2007lr &  0.1562 & 00 49 00.3 & +0 19 26.4 & high & $6.1 \times 10^{8}$ & $2.9 \times 10^{10}$ & $7.3 \times 10^{11}$  \\ 
2007ma &  0.1073 & 00 44 53.8 & +0 59 49.3 & high & $1.6 \times 10^{9}$ & $6.5 \times 10^{9}$ & $5.4 \times 10^{10}$  \\ 
2007mh &  0.1278 & 03 14 31.8 & +0 16 11.4 & high & $8.9 \times 10^{8}$ & 0 & $4.5 \times 10^{10}$  \\ 
2007mi &  0.1322 & 03 23 31.5 & +0 39 60.0 & low  & 0 & $1.6 \times 10^{10}$ & $5.2 \times 10^{11}$  \\ 
2007mj &  0.1232 & 03 34 44.4 & +0 21 19.9 & high & $2.5 \times 10^{8}$ & $2.5 \times 10^{10}$ & $1.5 \times 10^{11}$  \\ 
2007mm &  0.0665 & 01 05 46.7 & +0 45 31.8 & low  & 0 & $1.5 \times 10^{9}$ & $6.8 \times 10^{10}$  \\ 
2007mn &  0.0769 & 02 05 04.0 & +0 10 28.4 & high & $2.8 \times 10^{9}$ & $3.7 \times 10^{10}$ & $4.4 \times 10^{11}$  \\ 
2007nj &  0.1540 & 02 52 27.4 & +0 15 06.6 & low  & $4.7 \times 10^{9}$ & $1.7 \times 10^{10}$ & $3.8 \times 10^{11}$  \\ 
2007ok &  0.1655 & 02 28 24.3 & +0 11 04.8 & high & $3.0 \times 10^{8}$ & $3.6 \times 10^{10}$ & $4.3 \times 10^{11}$  \\ 
2007ol &  0.0560 & 01 37 23.7 & +0 18 43.2 & low  & 0 & $7.4 \times 10^{9}$ & $1.7 \times 10^{11}$  \\ 
2007om &  0.1052 & 23 54 20.7 & +0 55 03.4 & high & $3.9 \times 10^{8}$ & $7.9 \times 10^{10}$ & $7.6 \times 10^{11}$  \\ 
2007ou &  0.1132 & 02 23 42.7 & +0 49 33.6 & high & $2.2 \times 10^{9}$ & 0 & $1.2 \times 10^{11}$  \\ 
2007ph &  0.1294 & 20 51 13.4 & +0 57 20.9 & low  & 0 & $1.1 \times 10^{10}$ & $9.2 \times 10^{11}$  \\ 
2007pt &  0.1752 & 02 07 38.5 & +0 19 26.4 & high & $1.8 \times 10^{9}$ & 0 & $7.6 \times 10^{10}$  \\ 
2007px &  0.1080 & 00 22 44.0 & +0 28 44.4 & high & $2.6 \times 10^{10}$ & 0 & $5.3 \times 10^{11}$  \\ 
2007py &  0.2094 & 03 29 31.6 & +0 30 56.0 & low  & $1.6 \times 10^{9}$ & 0 & $1.3 \times 10^{11}$  \\ 
2007qa &  0.1085 & 01 52 33.9 & +1 14 38.7 & high & $1.6 \times 10^{9}$ & 0 & $1.3 \times 10^{11}$  \\ 
2007qr &  0.1359 & 02 52 29.2 & $-$1 08 22.3 & high & $6.0 \times 10^{8}$ & $1.4 \times 10^{9}$ & $5.7 \times 10^{10}$  \\
2007rs &  0.1241 & 00 46 27.4 & $-$1 03 44.1 & low  & 0 & $6.6 \times
10^{10}$ & $1.8 \times 10^{12}$  
\enddata
\tablenotetext{a}{Position of SN; data available from
  http://www.cfa.harvard.edu/iau/lists/Supernovae.html and
  http://www.cfa.harvard.edu/iau/cbat.html.}
\end{deluxetable*}%

\subsection{The High/Low Stretch Division} \label{sec:stretchdiv}

We wish to quantify the correlation between SN Ia stretch and host
galaxy stellar populations using the VESPA analysis.  Given our
limited sample size, we seek to split our SNe into two
populations based on their stretches.  
We weight each of our 101 supernovae by $M_{i, j}$, the stellar
mass in its host galaxy $j$ in a given VESPA age bin $i$.
We then define the total $M_i$ for supernovae with stretch $s$
lying in the interval $[a, a + \Delta s]$ (taking $\Delta s = 0.04$,
slightly larger than a typical error in the stretch parameter) as
\begin{equation}
M_i (s) \equiv \sum_{s_j \in [a, a + \Delta s]}
M_{i,j}
\end{equation}
with $j$ running over our 101 objects.  This quantifies the total star
formation rate in a given age bin associated with the SNe Ia in our
sample.  Note that $M_{i,j}$ is equal to the mass of {\it possible},
not actual, SN Ia progenitor systems.  Because galaxies are composed
of a mix of stellar populations of different ages, we cannot derive
the delay time distribution directly from these associated stellar
masses.  

\begin{figure}
\begin{center}
\includegraphics[width=\medfigwidth]{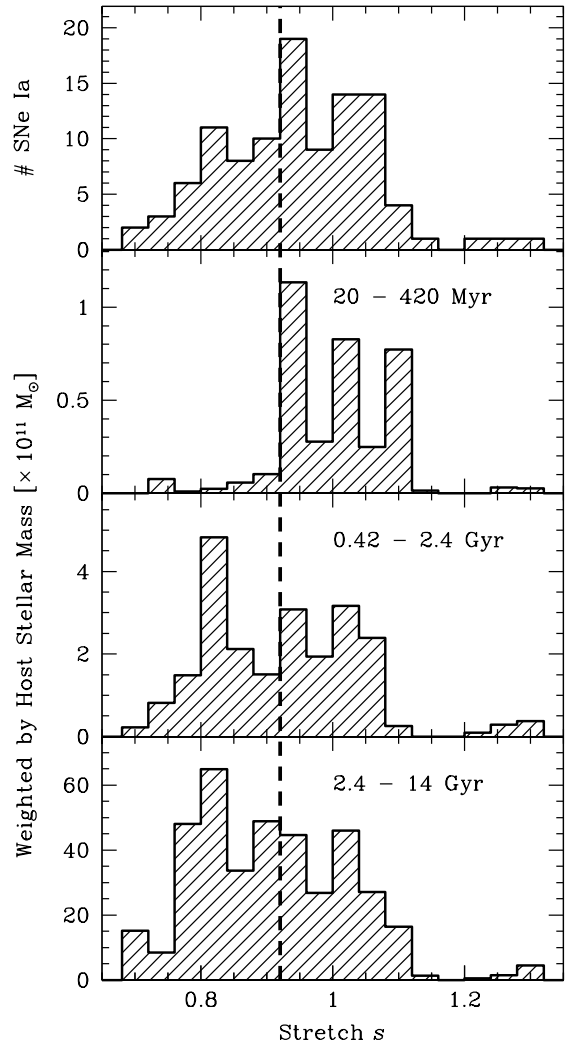}
\end{center}
\caption{Histograms of stellar mass of supernovae Ia hosts binned by
  SN stretch.  Objects in the top panel are unweighted, while SNe in
  the lower three panels are weighted by the stellar mass of their
  host galaxy in young (top), middle-aged (middle) and old (bottom)
  stars.  SNe with $s \gtrsim 0.92$ (to the right of the dotted line)
  are associated with young stars, while lower stretch SNe are not.
  To further explore this distinction, we define supernovae with $s >
  0.92$ as \emph{high stretch} SNe Ia. 
}
\label{fig:stretchcut}
\end{figure}

With this caveat in mind, the resulting distributions of stellar mass
as a function of stretch in different mass bins are shown in Figure
\ref{fig:stretchcut}.  The top panel is a histogram of stretches,
while the second, third, and fourth panels weight each SN by its
host stellar mass in young, middle, and old age bins, respectively.  
The second and fourth panels, in particular, show that lower
stretch SNe live in hosts with large quantities of old stars but little
recent star formation, while higher stretch SNe are in hosts with
abundant recent star formation.  Given the rough bimodality shown
here, we call Type Ia supernovae
\emph{high stretch} if they have $s > 0.92$, and \emph{low stretch}
otherwise.  In SDSS-SN, this division yields 60 high stretch and 41
low stretch SNe; the binnings are listed in Table \ref{tab:sne}. 

This result confirms the work of \cite{SullivanEtAl06},
\cite{NeillEtAl09}, \cite{Gallagheretal05}, and others:
that high stretch SNe are associated with star-forming galaxies.
We now seek to study this association in more detail and to derive a
quantitative delay time distribution for high and low stretch SNe Ia.

\subsection{Average Spectra} \label{sec:averagespectra}

The difference in stellar populations indicated by Figure
\ref{fig:stretchcut} should manifest itself in the spectra of the
hosts.  To examine this, we constructed average spectra of the high
stretch and low stretch hosts.  We corrected the spectra for Milky
Way extinction using the dust maps of
\cite{SchlegelEtAl98} and a \cite{Fitzpatrick99} $R_V = 3.1$
extinction law.  We scaled each spectrum to a common $r$ band fiber
magnitude, shifted the spectra to $z=0$, masked bad pixels flagged by
the SDSS pipeline, and weighted by the inverse variance in coadding.
To obtain robust estimates of the mean and variance of the average
spectra, we have used bootstrap resampling on the samples of 60 high
stretch and 41 low stretch hosts.  

\begin{figure}
\begin{center}
\includegraphics[width=\fullfigwidth,height=\fullfigheight]{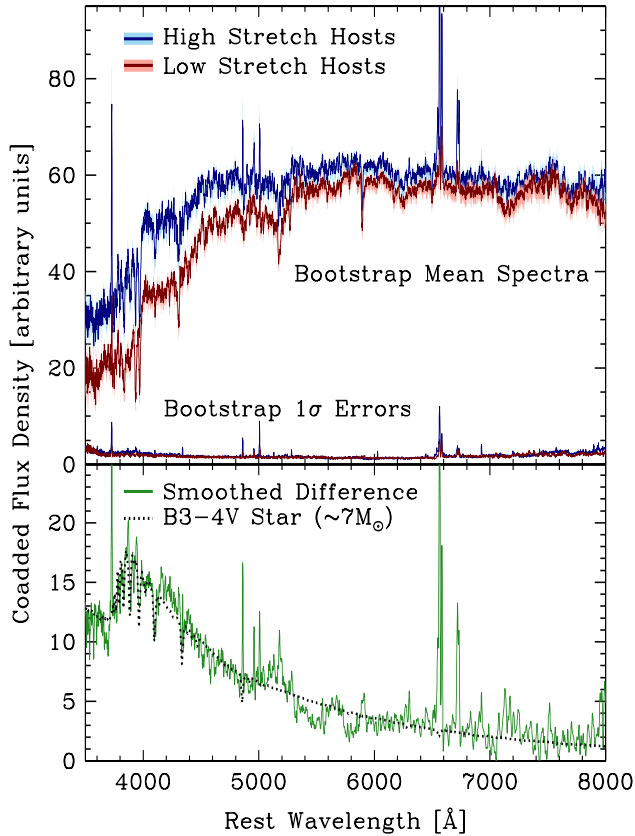}
\end{center}
\caption{Coadded SDSS spectra of hosts of 60 high stretch ($s > 0.92$
  - blue) and 41 low stretch ($s < 0.92$ - red) supernovae (upper
  panel); the shading shows the $1 \sigma$ error range derived from
  bootstrap resampling.  Note the strong blue continuum and nebular
  emission lines, indicative of recent and ongoing star formation, in
  the high stretch hosts.  The low stretch host spectrum has features
  characteristic of much older stellar populations.  The difference
  (lower panel) looks remarkably similar to the spectrum of a typical
  B star (note the strong Balmer series in absorption), suggesting
  young progenitors for high stretch Type Ia SNe.  The stellar
  spectrum is from \cite{1992ApJS...81..865S}.}
\label{fig:sdss_spectra}
\end{figure}

The results, shown in Figure
\ref{fig:sdss_spectra}, are striking.  The average spectrum of a
high stretch host shows exceptionally strong nebular emission lines
such as the Balmer series, $[\textsc{Oiii}]$, $[\textsc{Oii}]$,
$[\textsc{Nii}]$, and $[\textsc{Sii}]$ and a strong blue continuum, all
clearly indicative of recent and ongoing star formation.  The average
spectrum of low stretch supernova hosts, by contrast, shows absorption
lines characteristic of old stellar populations and little evidence of
interstellar gas.  The difference spectrum (lower panel) looks
much like the spectrum of a main-sequence B type star with nebular
emission lines superimposed.  Intriguingly, a single B3 spectrum from
\cite{1992ApJS...81..865S} fits the continuum and absorption lines of
the difference spectrum better than models of either a 50
Myr-old single stellar population, or 400 Myr of continuous star
formation.  While the bootstrap errors are too large to exclude either
of these possibilities, Figure \ref{fig:sdss_spectra} tantalizingly
suggests very young progenitors for some SNe Ia, as the difference
spectrum appears to require a population $\lesssim 50$ Myr old (B3
stars live for about 30 Myr).  Note that we cannot be absolutely
certain that the progenitors are this young, if bursts of star
formation typically last long enough so that a somewhat older
population is likely to be associated with ongoing star formation.  

Figure \ref{fig:sdss_spectra} confirms the difference in host galaxies
indicated in Figure \ref{fig:stretchcut} by the VESPA-derived star
formation histories.  High stretch, luminous Type Ia's are
associated with young O and B stars, while lower stretch SNe
are found in galaxies with much older stellar populations.  Figure
\ref{fig:sdss_spectra} offers a clear and dramatic confirmation of the
association between high stretch SNe and star formation.  The
average spectra change imperceptibly when the high/low stretch
boundary is varied from 0.90 to 0.92, representing a typical stretch
error as reported by SALT II.  Our results are therefore unlikely to
be affected by measurement errors in the stretch parameter.  

\section{Towards a DTD} \label{sec:constrain}

While Figure \ref{fig:sdss_spectra} is compelling on its own, we can
do better.  With a well-controlled survey, spectra for all galaxies in
the control sample, and VESPA star formation histories for all of
these galaxies, we now seek to calculate a delay time distribution in
the three age bins of Table \ref{tab:bin_info}.  We do not assume any
functional form for the efficiency of making SNe as a function of
progenitor age, but simply see what constraints the data alone 
can provide.  

Our method is as follows:  
\begin{enumerate}
\item Parametrize the DTD as an explosion rate per unit stellar mass
  $\epsilon_i$ for stars in each of three age bins $i$, 
\item Assume a prior probability distribution on the DTD explosion rates,
\item Select DTDs from this prior,
\item Generate samples of mock hosts from the DTDs and VESPA star
  formation histories of our control sample, 
\item Compare the average spectra of the mock hosts and observed hosts to
  calculate a likelihood for each DTD realization,
\item Repeat steps 3 - 5 many times to explore the $\epsilon_i$
  space and obtain a posterior probability distribution on the DTD.  
\end{enumerate}
We now discuss the general form of the delay time distribution and then
each of these steps.  

\subsection{The Delay Time Distribution}

Our star formation histories consist of the total stellar mass formed
in each of three age bins for each host galaxy.  The most general
delay time distribution is
therefore a set of three efficiencies, $\epsilon_i$, 
representing the mean number of SNe per unit stellar mass per year
from progenitors in age bin $i$ (Table \ref{tab:bin_info}).  We treat
high stretch and low stretch SNe separately in this analysis, giving
two sets of efficiencies $\epsilon_{h,i}$ and $\epsilon_{l,i}$.  

We may effectively remove two of these six parameters by requiring that
the sets of efficiencies be appropriately normalized, i.e.~that the
total number of expected SNe times their probability of detection
equal the number of SNe observed,
\begin{equation}
N_\mathit{SN, h} = \sum_{\mathrm{gals}~j} p_{h,j} (\mathit{detect}) t
\sum_i \epsilon_{h,i} M_{i,j},
\label{eq:snrate}
\end{equation}
where $p_{h,j}(\mathit{detect})$ is the probability that a supernova of
high stretch in galaxy $j$ enters the sample, $M_{i,j}$ is the amount
of stellar mass formed in age bin $i$ in galaxy $j$, $N_\mathit{SN,
  h}$ is the number of high stretch SNe Ia observed, and $t$ is the
duration of the survey.  An identical constraint applies to low
stretch supernovae.  

For future use, we also introduce normalized explosion efficiencies
corresponding to the fraction of high or low stretch supernovae
produced by stars in a given age bin, e.g.
\begin{equation}
\epsilon'_{h,i} \equiv \epsilon_{h,i} \frac{1}{N_\mathit{SN, h}}
\sum_{\mathrm{gals}~j} p_{h,j}(\mathit{detect}) M_{i,j}.
\label{eq:rescale}
\end{equation}
Thus, $\epsilon'_{h,i}$ and $\epsilon'_{l,i}$ each sum to unity.  

\subsection{Priors}

\begin{figure}
\begin{center}
\includegraphics[width=\figwidth]{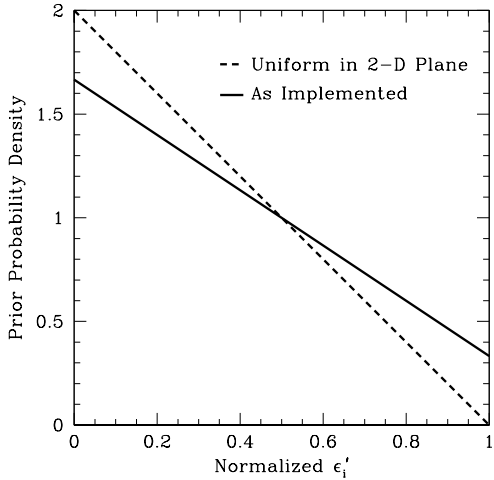}
\end{center}
\caption{Projected prior distributions for each $\epsilon'_i$, the
  fraction of high/low stretch SNe with progenitors in age bin $i$.
  The normalized $\epsilon'_i$ are related to the physical explosion
  rates $\epsilon_i$ by Equation \eqref{eq:rescale}.  A prior uniform in
  the 2-D plane defined by Equation \eqref{eq:snrate} would be
  strongly biased against DTDs concentrated in one age bin.}
\label{fig:priors}
\end{figure}

In order to use a Bayesian analysis to constrain the $\epsilon_{h,i}$
and $\epsilon_{l,i}$, we need to choose a prior distribution.  Following
\cite{LaPlace1824} and general practice in the literature, we seek a
uniform prior.  However, due to the normalization constraint, we cannot
place uniform priors simultaneously on all of the $\epsilon_{h,i}$ or
$\epsilon_{l,i}$.  An alternative would be to
choose our priors to be uniform in area over the two dimensional plane
defined by Equation \eqref{eq:snrate}.  However, the projection
of this prior onto any of the $\epsilon'_{h,i}$ yields a probability
distribution strongly peaked towards low values.  Physically, this
biases us against a DTD in which one age bin is responsible for a
large fraction of the SNe (see Figure \ref{fig:priors}).  
To avoid these combinatorial effects, we use a Monte Carlo
sampling that places a uniform prior on \emph{one} of the
$\epsilon'_{h,i}$ and \emph{one} of the $\epsilon'_{l,i}$ in each
realization.  By choosing $i$ randomly in each iteration, we retain
the symmetry between all of the $\epsilon'_{h,i}$ and
$\epsilon'_{l,i}$. 

We note that there are many possible priors that look qualitatively
similar to that shown in Figure \ref{fig:priors}.  We do not argue
that our choice is optimal, only that it is reasonable.  We have also
tested the dependence of our constraints on the choice of prior and
found little variation, as long the prior probability
density does not approach zero for any values of $\epsilon'_i$. 

\subsection{Creating a Mock Sample}

Given a model delay time distribution (i.e., sets of
$\epsilon_{h,i}$ and $\epsilon_{l,i}$),
we compute the expected number of high and low stretch supernovae
($\ll 1$) observed in each galaxy in the control sample.  For a galaxy
$j$, this is the sum of the explosion efficiencies times the mass in
each stellar age bin weighted by the detection probability:
\begin{equation}
\langle n_\mathit{SN,h,j} \rangle = p_{h,j}(\mathit{detect}) \sum_i
\epsilon_{h,i} M_{i,j}.
\label{eq:mean_sn}
\end{equation}
Our code thus generates realizations of $\{\epsilon'_i\}$ for both
high and low stretch, converts to
$\{\epsilon_i\}$ using Equation \eqref{eq:rescale}, and computes the
number of SNe Ia in each control galaxy as a Poisson random number
with a mean given by Equation \eqref{eq:mean_sn}. 

\begin{figure}
\begin{center}
\includegraphics[width=\figwidth]{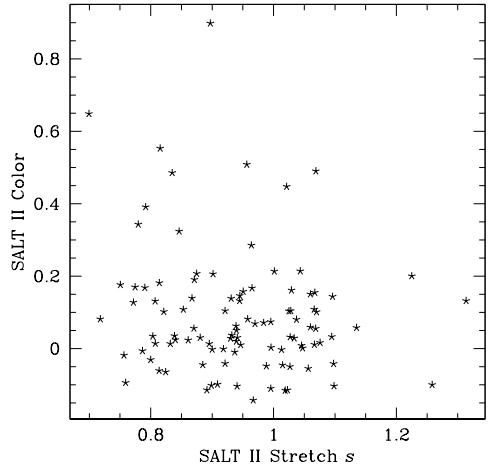}
\end{center}
\caption{Stretch vs. color ($B-V$ at peak $B$ magnitude relative to a
  fiducial average), as determined by SALT II.  The two quantities
  are not correlated, which allows us to fold the color variation into
  the detection efficiency (Equation \ref{eq:detect}).}
\label{fig:stretch_color}
\end{figure}

For the detection probability, we use a crude estimate of the
selection function and tweak the parameters to match the approximate
redshift range of the subset of SDSS-SN with host galaxy spectra.  We
adopt the functional form 
\begin{equation}
p(\mathit{detect}) = \left[1 + \mathrm{exp}\left(\frac{m_\mathit{peak}
  - m_\mathit{lim}}{a} \right) \right]^{-1},
\label{eq:detect}
\end{equation}
where $m_\mathit{peak}$ is the supernova's
peak SDSS $r$ band magnitude, $m_\mathit{lim}$ is an approximate
limiting magnitude of the survey, and $a$ is a softening parameter to
account for the dispersion in SN peak magnitudes and colors and the
survey detection efficiency.  Because this model does not naturally
capture the fact that higher stretch SNe are above any threshold for a
longer period of time, we allow the parameters $m_\mathit{lim}$ and
$a$ to differ for high and low stretch objects.  Fitting to the
observed redshift distribution, we adopt $m_\mathit{lim} = 19.5$ and
$a = 0.6$ for low stretch SNe, and $m_\mathit{lim} = 20.2$ and $a =
0.4$ for high stretch SNe.  The peak supernova magnitude in $r$ is
given by 
\begin{equation}
m_\mathit{peak} = M_\mathit{stand} - \alpha (s - 1) + \mu(z) +
K(s, z)+A_r
\label{eq:mag}
\end{equation}
where $M_\mathit{stand}$ is the standardized peak B absolute magnitude,
$\alpha$ is the \cite{Phillips93} parameter relating stretch and peak
luminosity, $s$ is a typical stretch,
$\mu(z)$ is the distance modulus, $A_r$ is the $r$ band Milky Way
extinction along the supernova's line of sight \citep{SchlegelEtAl98},
and $K(s,z)$ is an approximate \emph{K}-correction to the SDSS $r$
band at $z=0$ (see \cite{2002PASP..114..803N} for details on the
$K$-correction).  
We adopt the values $M_\mathit{stand} = -19.41$ and $\alpha = 1.56$
\citep{GuyEtAl05}, and take the typical values of $s$ to be
our sample medians, $1.02$ for high stretch and $0.83$ for low stretch
SNe Ia.  Since we fit for $m_\mathit{lim}$ and $a$ to
reproduce the observed redshift distribution {\it after} fixing all other
parameters, changing the standardized absolute magnitude $M_\mathit{peak}$
will not affect our results.  Because we fit high and low stretch SNe
separately, the value of $\alpha$ similarly has no effect.  Equation
\eqref{eq:mag} does not account for variation in
color ($B-V$ at peak $B$ magnitude), which is instead absorbed into
the softening parameter $a$.  If color is correlated with stretch,
Equation \eqref{eq:mag} will not fully capture differences between the
detectability of high and low stretch SNe.  However, as shown in
Figure \ref{fig:stretch_color}, we find no evidence of such a
correlation.  

\begin{figure}
\begin{center}
\includegraphics[width=\medfigwidth]{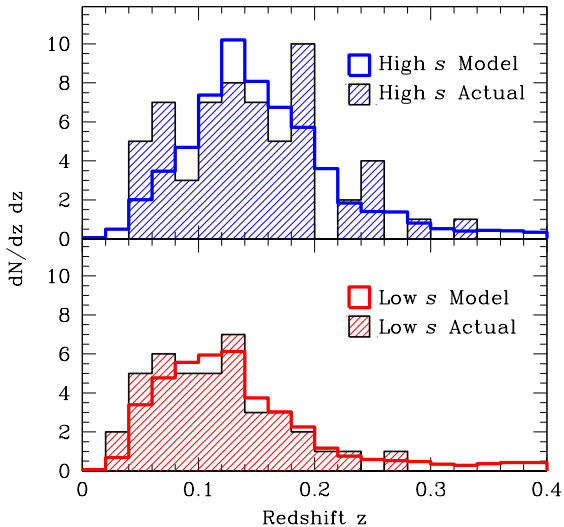}
\end{center}
\caption{Redshift histograms of the observed SNe with SDSS host galaxy
  spectra against the histograms of one realization of mock hosts with
  the detection probability given by Equation \eqref{eq:detect}.  We
  adopt values of $m_\mathit{lim} = 20.2$ and $a = 0.4$ for high
  stretch SNe, $m_\mathit{lim} = 19.5$ and $a = 0.6$ for low stretch
  SNe.}
\label{fig:zhist}
\end{figure}

Figure \ref{fig:zhist} compares the redshift distribution of a mock
sample generated by Equations \eqref{eq:mean_sn}, \eqref{eq:detect},
and \eqref{eq:mag} with the observed redshift distribution.  While
there are slight differences, the means and widths of the
distributions agree reasonably well for both high and low stretch SNe.
The detection probability given by Equation \eqref{eq:detect} is
intended only to be approximate and to set the redshift range of
galaxies that serve as potential hosts.  Modest variations in
Equations \eqref{eq:detect} and \eqref{eq:mag} have almost no effect
on our results, particularly because we compare spectra weighted by
their inverse variances: the slight discrepancy at high redshift is
dominated by galaxies with poor signal-to-noise ratio spectra that
have a very small impact on the average spectrum.  As an additional
check, we have verified that Equations \eqref{eq:mean_sn},
\eqref{eq:detect}, and \eqref{eq:mag} successfully reproduce the
observed range of host galaxy masses as derived from VESPA.  

\subsection{A Likelihood Function}

To use Bayes' Theorem, we need to calculate the likelihood of observed
host properties given a sample of mock hosts selected with a specific
model for the $\{\epsilon_i\}$.  We use a comparison metric based on
the average spectra of the host galaxies; essentially, with our Monte
Carlo realizations of delay time distributions, we seek to reproduce
the mean spectra shown in Figure \ref{fig:sdss_spectra}.  We use a
$\chi^2$ statistic to compare the average spectra of our Monte Carlo
realizations to those of the observed hosts. 

To obtain useful constraints, we
need both very high signal-to-noise ratio average spectra for the mock
samples, and reliable estimates of the mean and variance associated
with the average spectra of the observed hosts.  The first requirement
is met by using a large mock sample drawn from our 77,000 control
galaxies.  The second problem is solved using bootstrap resampling of
the 60 high stretch and 41 low stretch actual hosts, as described in
\S\ref{sec:averagespectra} and shown in Figure \ref{fig:sdss_spectra}.
As a check, we compared the bootstrap errors to the error-weighted
pixel-by-pixel scatter between spectra and found very good agreement.  

With an average spectrum and bootstrap errors for the SN hosts,
we may assign a $\chi^2$ goodness-of-fit value to the mean spectrum of
the mock hosts by averaging the $\chi^2$ values over all pixels.  We
then use this reduced $\chi^2$ to compute a rejection probability.  

There are two reasons why we have not directly compared
the stellar populations of SN Ia hosts and non-hosts.  First, such a
comparison would require us to apply VESPA to the SN hosts to derive
their star formation histories.  We are limited by $\sqrt{N}$
statistics in the hosts, and errors in the VESPA outputs are much
larger than in the spectra.  Second, the recovered stellar masses
would be used twice for the control galaxies:  once to determine how
many supernovae to set off in the mock samples (Equation
\ref{eq:mean_sn}), and once to compare the resulting mock hosts to the
observed hosts. Any metric would therefore contain the sum of the
squares of the recovered stellar masses, so that random variances in
these masses would {\it add} and produce a bias.  Our chosen metric,
by combining spectra rather than derived quantities, avoids both
drawbacks.  

\section{Results} \label{sec:results}

We have run 100,000 Monte Carlo realizations of the delay time
distribution from the prior shown in Figure \ref{fig:priors}.  In
each realization, we draw a set of normalized explosion efficiencies
($\epsilon'$, Equation \ref{eq:rescale}), convert to the physical
explosion efficiencies ($\epsilon$, Equation \ref{eq:snrate}), and use
Equations \eqref{eq:mean_sn}, \eqref{eq:detect}, and \eqref{eq:mag} to
generate a sample of mock hosts.  We then
construct the composite spectra for the mock high stretch and low
stretch hosts and compare them to the spectra of observed hosts shown
in Figure \ref{fig:sdss_spectra} using a $\chi^2$ test.  We take the 
likelihood of each realization to be the rejection probability
($Q$ function) of its computed value of $\chi^2$.  In
practice, our best-fit models do produce formally good fits, with
$\chi^2$ per pixel typically in the range 0.65 - 1.05.  Normalizing these
likelihoods by their sum over all Monte Carlo realizations, we obtain
the posterior probability distributions for the explosion
efficiencies.  The distribution for a single efficiency is computed by
integrating over the other variables.  

\begin{figure}
\begin{center}
\includegraphics[width=\figwidth]{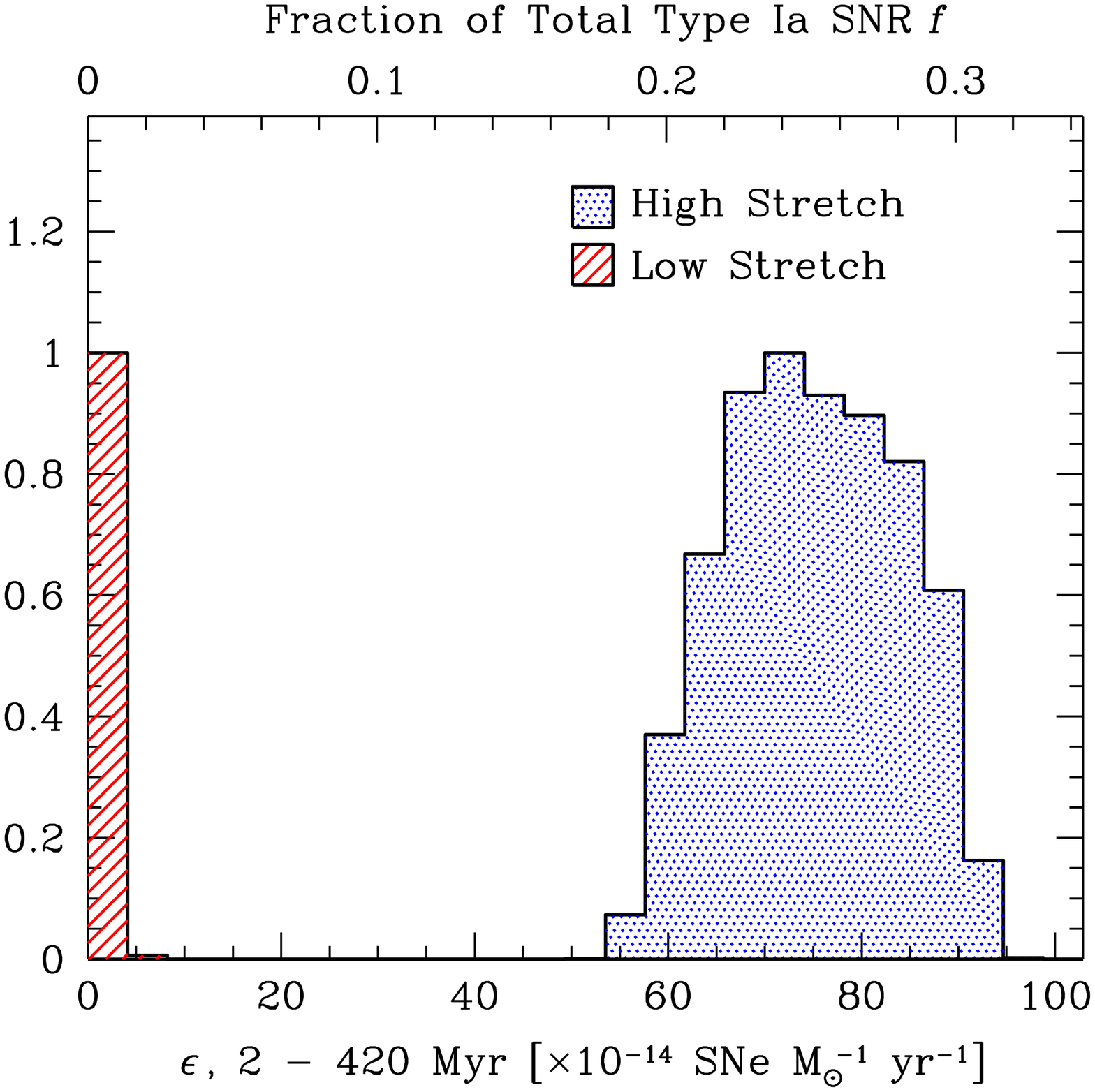} 
\includegraphics[width=\figwidth]{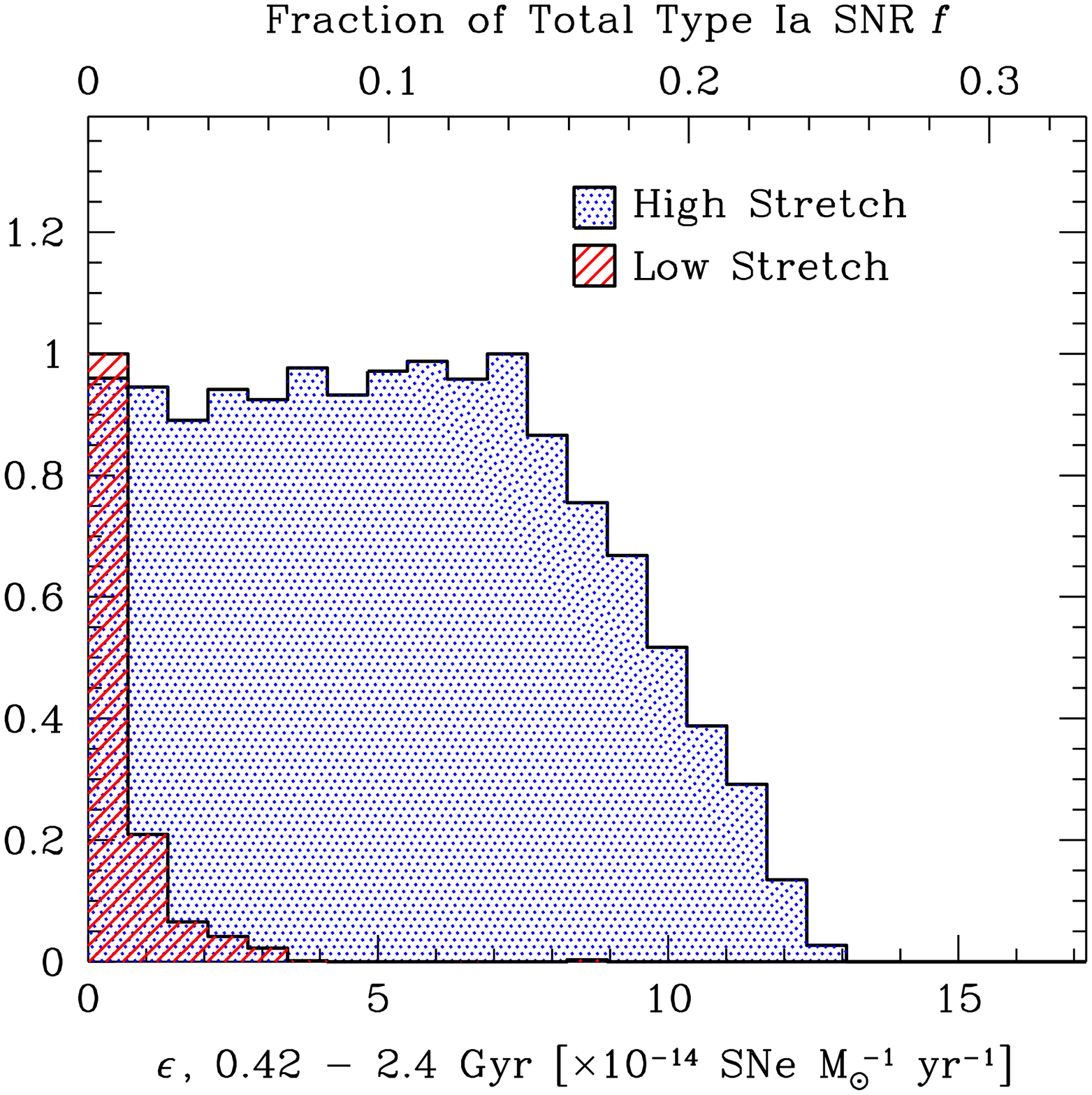} 
\includegraphics[width=\figwidth]{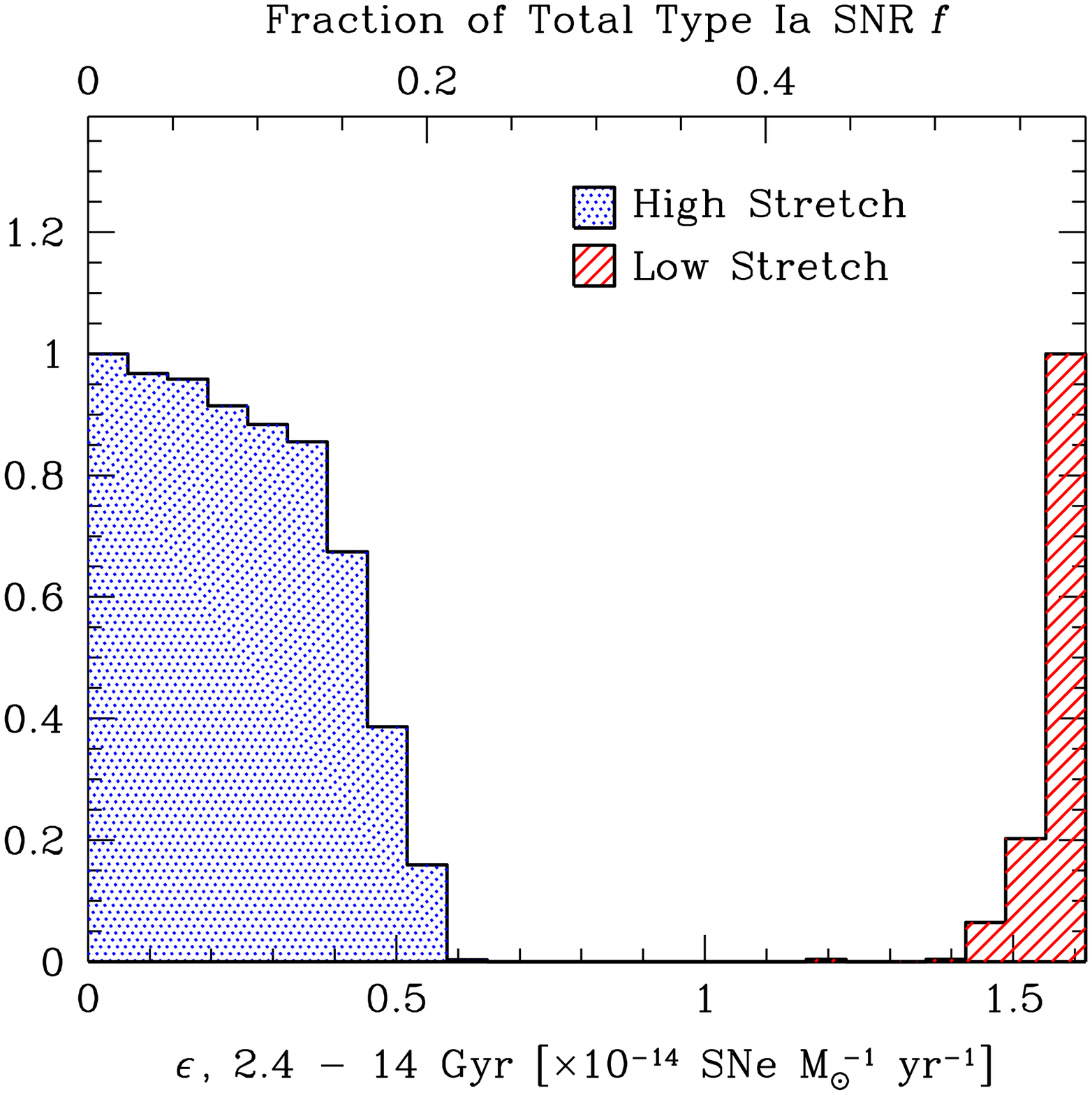}
\end{center}
\caption{The posterior probability distributions of the DTD.  The
  lower horizontal axes give the explosion efficiencies
  $\epsilon_{h,i}$ and $\epsilon_{l,i}$, the supernova rates per unit
  stellar mass per year for progenitors in age bin $i$ (Equation
  \ref{eq:snrate}).  The upper axes show $f_{h,i}$ and $f_{l,i}$, the
  proportions of the current observed SN Ia rate in the SDSS Stripe 82
  galaxies (Equation \ref{eq:fractions}).  Nearly all high
  stretch SNe have progenitors $\lesssim 400$ Myr old (top panel), while 
  Figure \ref{fig:sdss_spectra} suggests that the typical age may be even
  younger.  Low stretch SNe have a characteristic delay time of at
  least 2-3 Gyr (bottom panel), with essentially no contribution from young
  progenitors.  } 
\label{fig:results}
\end{figure}

The resulting probability distributions are shown in Figure
\ref{fig:results}.  The lower horizontal axes give the explosion
efficiencies in physical units ($\epsilon_{h,i}$ and $\epsilon_{l,i}$,
Equation \ref{eq:snrate}), while the upper axes show the fraction
of all SNe Ia in the SDSS spectroscopic sample formed as high or low
stretch from a given progenitor age bin.  These are computed from the
$\epsilon_{h,i}$ or $\epsilon_{l,i}$ as, e.g.,
\begin{equation}
f_{h,i} = \frac{\epsilon_{h,i}}{N_{\mathit{SN}}}
 \sum_{\mathrm{gals}~j} p_{h,j} (\mathit{detect}) M_{i,j},
\label{eq:fractions}
\end{equation}
where $p_{h,j} (\mathit{detect})$ is the detection probability for
high stretch SNe in galaxy $j$, $M_{i,j}$ is the stellar mass formed
in age bin $i$ in galaxy $j$, $N_{\mathit{SN}}$ is the total (high and
low stretch) number of SNe detected, and the sum is taken over control
galaxies.  Equation \eqref{eq:fractions} differs from Equation
\eqref{eq:rescale} only in the denominator: while $\epsilon'_{h,i}$
and $\epsilon'_{l,i}$ each sum to unity, $f_{h,i}$ and $f_{l,i}$ {\it
  together} sum to unity.  Note that because a volume-limited sample
of galaxies will have different properties from the SDSS spectroscopic
sample (in particular lower masses), $f_{h,i}$ or $f_{l,i}$ for the
full sample of SDSS supernovae may differ from the values we recover.

\begin{figure}
\begin{center}
\includegraphics[width=\figwidth]{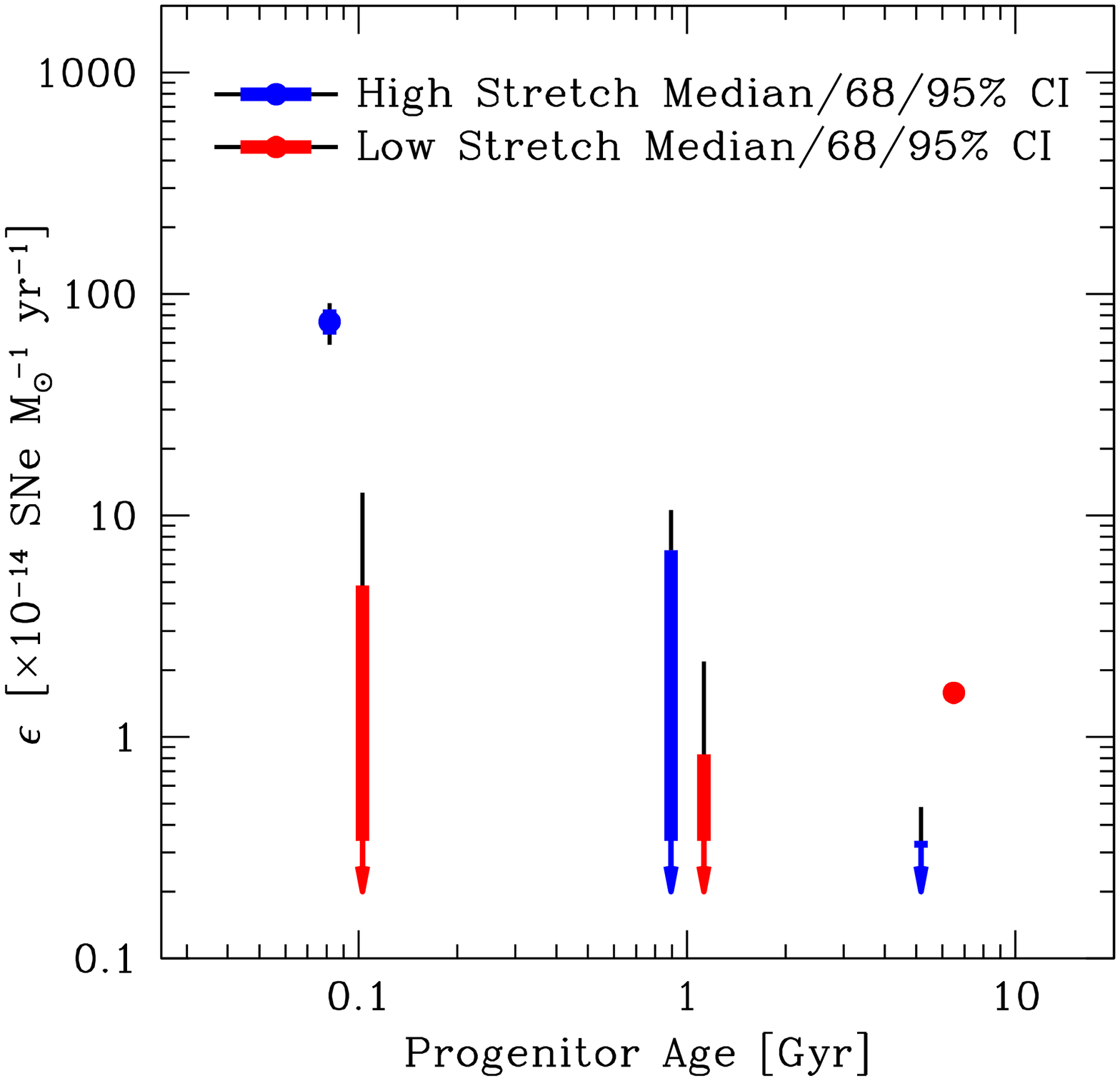}
\includegraphics[width=\figwidth]{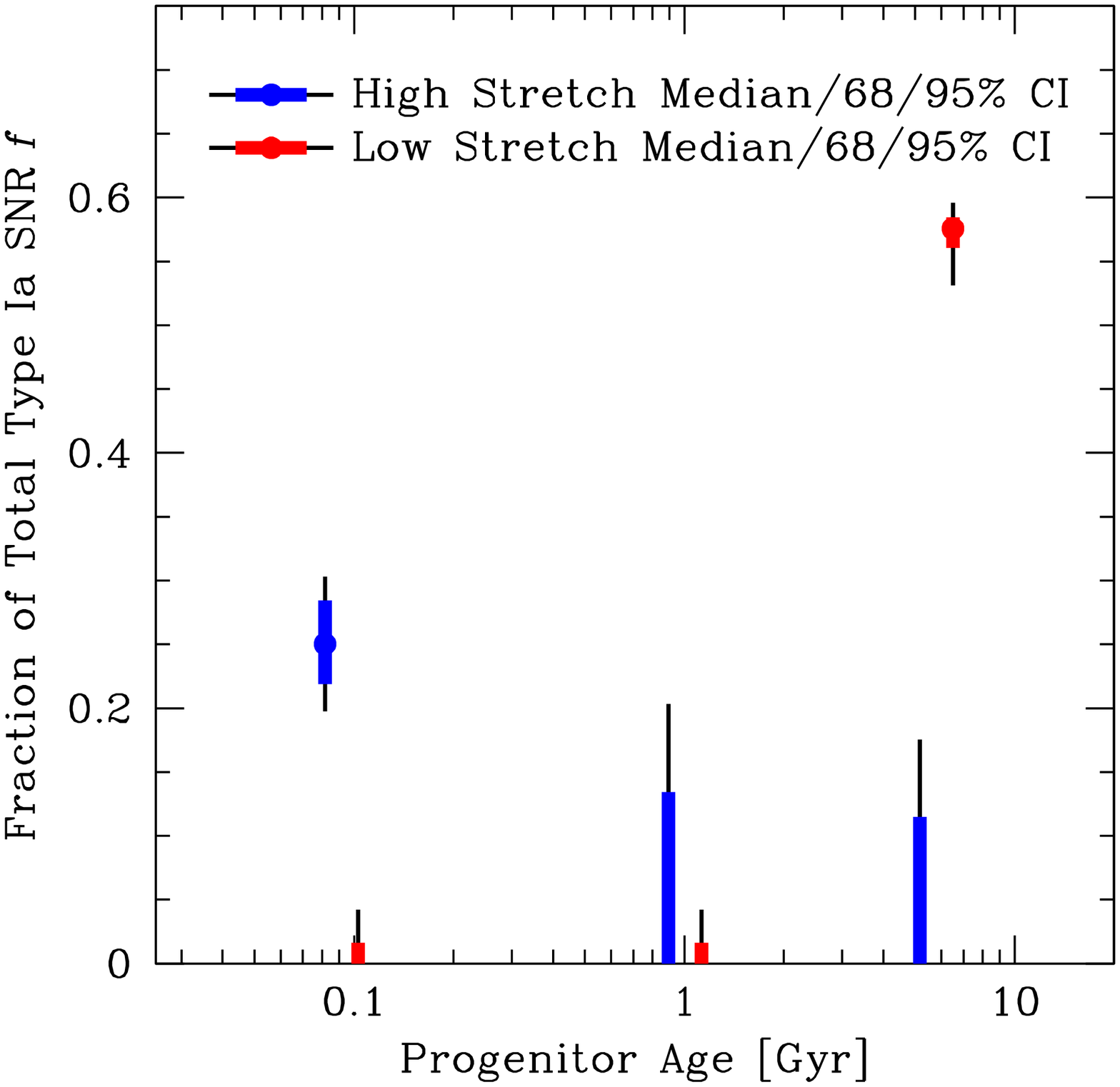}
\end{center}
\caption{The recovered DTDs, expressed in physical units (Equation
  \ref{eq:snrate}, top panel) and as a proportion of the total
  SN Ia rate (Equation \ref{eq:fractions}, bottom panel) in
  the SDSS spectroscopic sample.  Upper limits are
  shown for distributions peaking near zero, medians and two-sided
  confidence intervals are shown otherwise.  Points are drawn at the
  midpoint of their relevant age ranges (see Table
  \ref{tab:bin_info}), and the high and low stretch bars are offset
  for ease of viewing.}
\label{fig:dtd_results}
\end{figure}

These same quantities, $f$ and $\epsilon$, are also plotted in Figure
\ref{fig:dtd_results}, with the top panel showing the supernova rate
in physical units ($\epsilon_{h,i}$ in blue and $\epsilon_{l,i}$ in
red) and the bottom panel showing the fraction of SNe from a given
range of progenitor ages ($f_{h,i}$ in blue and $f_{l,i}$ in red).
The dots show the median values, while the colored and black bars
represent the $68\%$ and $95\%$ confidence intervals, respectively.
For distributions peaked near zero, only upper limits and one-sided
confidence intervals are shown.  The constraints on $f$ and $\epsilon$
are also listed in Table \ref{tab:results}.  

The results are striking.  We can constrain most high stretch
SNe Ia to have progenitors younger than $\lesssim 400$ Myr, and Figure
\ref{fig:sdss_spectra} tantalizingly suggests an even younger
characteristic age (B3 stars live for $\sim$30 Myr).  
While young stars dominate the production of high stretch SNe Ia, they
make no significant contribution to the low stretch rate.  Instead,
these supernovae have a characteristic delay time of at least 2-3
Gyr.  Intermediate progenitors, evolving on the time scale of a Main
Sequence A star, contribute little to either the high or low stretch
rate.  In the SDSS Stripe 82 galaxies, the prompt, high
stretch channel and the delayed, low stretch channel each account for
roughly half of all Type Ia supernovae.  

The formal constraints we obtain for the low stretch rate from young
stars (Figure \ref{fig:results}) are especially stringent.  Because
even blue galaxies have lots of old stars, we find it difficult to
reproduce a spectrum as quiescent as that shown in Figure
\ref{fig:sdss_spectra}:  a sample selected purely by old stellar mass
will still include star-forming galaxies.  Our measured spectra are
{\it only} consistent with a delay time distribution in which all (or
nearly all) low stretch SNe Ia are produced by old stars.
Nevertheless, we cannot rule out any progenitor channel expected to
produce $\lesssim 1$ detectable supernova over the entire population
of SDSS galaxies, and we have therefore set a floor on our SN 
rate constraints from Poisson statistics.  This floor determines the
limits on $\epsilon_{l,1}$ and $\epsilon_{l,2}$, the low stretch
rates from young and middle-aged stars, in Table \ref{tab:results} and
Figure \ref{fig:dtd_results}.  

To compare our results with earlier work based on the $A+B$ model
(Equation \ref{eq:AB_model}), we need to interpret $A$ and $B$ in our
framework.  We take $A$, the coefficient of the total
stellar mass, to be our supernova rate for old stars, $\epsilon_3$.
We convert $B$, the coefficient of current star formation (in
$\mathrm{SNe}\, M_\odot^{-1}$), to our rate for young stars
$\epsilon_1$ (in $\mathrm{SNe}\,M_\odot^{-1}\,\mathrm{yr}^{-1}$) by
multiplying by the temporal width of our first age bin, about 400 Myr.  
Our results agree reasonably well with those of \cite{NeillEtAl06}
(young $\epsilon_1 = 200 \pm 50$, old $\epsilon_3 = 1.4 \pm 1.0$ in our
units of $10^{-14}~\mathrm{SNe}\,M_\odot^{-1}\,\mathrm{yr}^{-1}$), but
less well with those of \cite{SullivanEtAl06} ($\epsilon_1 = 100 \pm
20$, $\epsilon_3 = 5.3 \pm 1.0$) or \cite{MannucciEtAl05} ($\epsilon_1
= 650 \pm 280$ or $300^{+180}_{-150}$, $\epsilon_3 =
4.4^{+1.6}_{-1.4}$).  Much of the discrepancy is due to the definition
of stellar mass:  \citeauthor{SullivanEtAl06} and
\citeauthor{MannucciEtAl05} apply a correction for dead stars,
decreasing stellar mass and correspondingly increasing $\epsilon$.
\cite{DildayEtAl08}, who measured these parameters using SDSS-SN
without correcting for dead stars, obtained $\epsilon_1 =
230^{+85}_{-78}$ and $\epsilon_3 = 2.8 \pm 1.2$ .  In addition, the
values of $\epsilon_i$ depend on the authors' choice of stellar models
and proxies for recent star formation, and may therefore
be expected to disagree.  Further, while our relative rates for
progenitors of different ages are extremely robust to the details of
the detection function (Equation \ref{eq:detect}), the normalization
of our rates is less so.  A poor estimate of the detection function
would multiply all of the rates $\epsilon_{h,i}$ and $\epsilon_{l,i}$
in Table \ref{tab:results} by a constant of order unity.  

\begin{deluxetable}{ccccc}

\tablecaption{TABLE \ref{tab:results}\\The DTD: results of the Monte Carlo}
\tablewidth{0pt}
\tablehead{
\colhead{Age Range (Gyr)} & \colhead{$\epsilon_{h}$\tablenotemark{a}} &
\colhead{$f_{h}$\tablenotemark{b}} &
\colhead{$\epsilon_{l}$\tablenotemark{a}} &
\colhead{$f_{l}$\tablenotemark{b}}
}
\startdata
0.002 -- 0.42 & $75^{+16}_{-16}$ & $0.25^{+0.05}_{-0.05}$ & $< 13$ &
$< 0.04$ \\ 
0.42 -- 2.4   & $< 13$ & $< 0.20$ & $< 2.2$ & $< 0.04$ \\
2.4 -- 14     & $< 0.48$ & $< 0.18$ & $1.58^{+0.06}_{-0.12}$ &
$0.58^{+0.02}_{-0.03}$ 
\enddata
\tablenotetext{a}{Defined in Equation \eqref{eq:snrate}, units are
  $10^{-14}~\mathrm{SNe}\,M_\odot^{-1}\,\mathrm{yr}^{-1}$.}
\tablenotetext{b}{Defined in Equation \eqref{eq:fractions}.}
\tablecomments{The errors given are 95\% confidence intervals.}
\label{tab:results}
\end{deluxetable}%

\section{Discussion and Conclusions} \label{sec:discussion}

\subsection{Robustness and Systematics}

SDSS-SN is a large, controlled, untargeted survey, and is therefore
largely free of systematics in target selection.  The subset with host
galaxy spectra does have additional selection criteria that can lead
to systematics; however, the host galaxies were selected from the
entire Stripe 82 spectroscopic sample only through the occurrence of a
detectable SN Ia.  Further, our comparison of Monte Carlo and observed
host galaxies is based not on derived quantities (like stellar
masses), but rather on the spectra themselves.  We therefore believe
our results to be robust.  Here, we address several possible sources
of bias and argue that all should be minor.  

One possible source of systematics is VESPA and its input stellar
models (see \S\ref{sec:vespa}).  We have chosen our models and the
resolution of our recovered star formation histories to minimize these
effects.  More importantly, we have avoided any use of VESPA on the
actual hosts (other than in Figure \ref{fig:stretchcut}, which we use
solely to set the stage), and only use it to select our samples of
mock hosts.  Therefore, the star formation histories from VESPA will
only bias our results if they are incorrect in the {\it mean}.  The
clear evidence of young stars in Figure \ref{fig:sdss_spectra}
qualitatively supports the results in Figure \ref{fig:results}.  

Our earlier paper, \cite{AubourgEtAl08}, used VESPA to show the
existence of a prompt component associated with the age bin to 180
Myr.  \citeauthor{AubourgEtAl08} and the present manuscript use
different approaches to recover information about SN Ia progenitors.
While in this paper we aim at recovering the full DTD in the least
parametric possible way allowed by the data, in
\citeauthor{AubourgEtAl08} we focused on recovering the shortest age
of the prompt progenitor channel.  It is possible that very young star
formation can mask the presence of older stars.  This in itself does
not remove the requirement for the presence of a young population, as
implied by the analysis of \citeauthor{AubourgEtAl08}, but in order to
avoid this issue, we have taken a cautious approach and combined the
young bins into a broad bin.  However, the difference spectrum in
Figure \ref{fig:sdss_spectra} suggests that the youngest bin in our
study may be dominated by progenitors substantially younger than 400
Myr, which would give support to the \citeauthor{AubourgEtAl08}
findings.  Note that we cannot with these data exclude the possibility
that the progenitors are older, as we have demonstrated only that very
recent star formation is required to be present.  An older progenitor
population could be reconciled provided correlations in the star
formation rate mean that older star formation is, in the majority of
cases, accompanied by a very young population.  We are limited by the
signal-to-noise ratio of the difference spectrum; with a larger
supernova host sample, we should be able to model better the age of
the stellar populations of the difference spectrum.  

Systematic errors in the shape of the observed spectra will give
errors in the VESPA-derived star formation histories.  In a sample of
physically identical galaxies, such an effect could, for example,
produce a subsample with bluer measured spectra and thus younger
inferred stellar populations.  Selecting galaxies based on their
derived masses of young stars will therefore produce average spectra
that are too blue.  In this way, differences in the average spectra of
mock hosts selected by Equation \eqref{eq:mean_sn} will be exaggerated
by using the VESPA-derived masses rather than the (unknown) physical
masses.  However, this effect would make it {\it easier} to reproduce
the differences in average spectra seen in Figure
\ref{fig:sdss_spectra}, biasing us {\it against} the different
progenitor ages for high and low stretch SNe Ia seen in Figures
\ref{fig:results} and \ref{fig:dtd_results}.  These effects are also
small: \cite{Adelman-McCarthy08} show that spectrophotometric
calibrations are accurate to about 4\% rms (see their Figures 4 and
5).  

Another possible source of systematics is the variation in SN and host
galaxy properties.  Galaxies with young stellar populations tend to
have more dust, which could render SNe associated with younger stellar
populations fainter.  We have not taken host galaxy dust into account
other than by adding scatter to our selection function, through the
$a$ parameter in Equation \eqref{eq:detect}.  However, the colors of
SNe are measurable and depend in part on host galaxy extinction.  As
shown in Figure \ref{fig:stretch_color}, the color parameter recovered
by SALT II is uncorrelated with the stretch.  Perhaps more
surprisingly, it is also uncorrelated with the mean interstellar dust
extinction as fit by VESPA.  Any systematic variations of color with
host galaxy properties must therefore be small, and should have little
effect on our results.  

\cite{KellyEtAl09} have found a hint of another systematic effect, a
correlation between peak SN Ia luminosity and host galaxy mass not
captured by the color or stretch variation.  However, the magnitude of
this effect is too small ($\lesssim 0.1$ magnitude) to significantly
impact our study.  

In addition to these systematics, we could suffer biases either from
our light curve ratings (\S\ref{sec:stretch}) or from our choice of
parameters in Equations \eqref{eq:detect} and \eqref{eq:mag}.  To
eliminate the former, we did all of the light curve fits and ratings
blindly (\S\ref{sec:stretch}), with no knowledge of the host galaxies.
For the latter, we took all free parameters in Equation \eqref{eq:mag}
from \cite{GuyEtAl05} and fit the parameters $m_\mathit{lim}$ and $a$
in our detection efficiency (Equation \ref{eq:detect}) to match the
observed redshift distributions of both high and low stretch SNe.
Changing $m_\mathit{lim}$ and $a$ could multiply all of the recovered
explosion efficiencies $\epsilon$ by a constant of order unity, but
our results are otherwise insensitive to the details of the selection
function. 

Finally, we measure the SN Ia rates in the SDSS spectroscopic sample,
which contains fewer low-mass, metal-poor galaxies than a
volume-limited sample would.  Should SN Ia production be suppressed in
low metallicity environments as \cite{KobayashiEtAl98} and others have
suggested, SN Ia rates could be lower in a volume-limited sample than
our results would indicate.  

\subsection{Connection to Progenitor Models}

It is remarkable how cleanly our observational cut in stretch divides
the sample into groups with distinct progenitors.  The choice of
$0.92$ as a division is somewhat arbitrary; it was a guess motivated
by Figure \ref{fig:stretchcut} and \cite{SullivanEtAl06}.  The
observed distribution of stretches does not show a clear bimodality
(see Figure \ref{fig:stretch_color}) and indeed appears continuous.
It is of course possible that SNe Ia have {\it more} than two
progenitor channels, each distinguished by a range of stretches.
However, Figure \ref{fig:stretchcut} provides little guidance on where
to split either the high or low stretch sample, and in any case, our
sample sizes of 60 high stretch and 41 low stretch SNe are too small
to profitably subdivide further. 

Given the division by stretch into two samples, we can test the
compatibility of our DTD with predictions of various progenitor
models.  We find that many can explain the prompt channel but have
extreme difficulty reproducing our low stretch DTD, in which nearly
all systems take more than $\sim$$2.4$ Gyr to explode.  The
theoretical single degenerate DTD calculated by \cite{WangEtAl10} is
compatible with our high stretch sample but falls short of our low
stretch rates by at least an order of magnitude (see their Figures 8
and 9).  All of the DTDs published by \cite{Greggio05} have too high a
rate from young stars to match our low stretch sample, though several
are compatible with our high stretch rates.  \cite{RuiterEtAl09}
calculate two-peaked theoretical DTDs for various common envelope
scenarios, but with the exception of one semidetached double white
dwarf binary model, find rates in young stars inconsistent with our
low stretch constraints.  It is possible that a single progenitor
channel could produce luminous, high stretch objects with a short
delay time and subluminous, low stretch objects with a long delay
time, perhaps because of different compositions at white dwarf birth.
Though recent progress has been made \citep{WoosleyEtAl07}, we still
lack an understanding of the dependence of SN Ia stretch on the
progenitor composition, and do not attempt to address the likelihood
of this scenario in this paper.  

\subsection{Conclusions}

We have used SDSS-SN, the only large untargeted supernova survey at
low redshift, to constrain the progenitor populations of Type Ia
supernovae.  The blind nature of the survey renders it free of most
systematics, while our use of spectra of both hosts and a large
control sample has allowed us to minimize the impact of stellar
population models on our results.  We dramatically confirm the two
populations seen by \cite{SullivanEtAl06}, \cite{MannucciEtAl06},
\cite{ScannapiecoBildsten05}, \cite{AubourgEtAl08} and others, finding
a ``prompt'' component of luminous, high stretch SNe with a
characteristic delay time $\lesssim 400$ Myr (the difference spectrum
in Figure \ref{fig:sdss_spectra} hints at a time as short as tens of
Myr) and a ``delayed'' component of subluminous, low stretch SNe with
a delay time $\gtrsim 2.4$ Gyr.  While our results are in broad
agreement with the $A+B$ model (Equation \ref{eq:AB_model}), they
place strong constraints on the progenitor ages and cause difficulties
for many extant theoretical delay time distributions.  We caution
against any physical interpretation of the precise values of the age
boundaries, which serve only as limits of integration.  

Type Ia supernovae have played a key role in our current understanding
of the cosmological model.  In spite of our incomplete understanding
of Type Ia progenitors and explosions, SN Ia surveys like SNLS
\citep{AstierEtAl06}, ESSENCE \citep{Wood-VaseyEtAl07} and
SDSS-SN \citep{KesslerEtAl09} continue to provide the best
constraints on cosmological parameters.  Further improvements, for
example with the JDEM candidate SuperNova Acceleration Probe
\citep{SNAP_collab04}, will require systematic effects in SN
Ia standardization to be controlled to $1-2\%$.  The identification of
the stellar evolution paths that can yield SNe Ia is therefore a key
issue: the brightness of a supernova could depend on the nature
of its progenitor, and the demographics of SNe Ia would depend on
redshift through the evolution time.  This could yield an effective,
non-cosmological, dependence of the mean ``standardized'' absolute
magnitude on redshift that would bias dark energy measurements. In
principle such an effect could be advantageous, as analyses of host
spectra could be used to determine the probabilities of each
progenitor route, and hence the use of the appropriate
\cite{Phillips93} stretch-luminosity relation.  This will require both
better constraints on Type Ia progenitor models and an improved
understanding of the Phillips relation.  

While Type Ia supernovae are most widely studied because of their use
as standard candles, they are also dynamically important in the
interstellar medium and are believed to be the main source of
iron-peak elements.  A fast route to Type Ia supernovae therefore has
implications for the interpretation of alpha-enhancement.  If most SNe
Ia are old, iron enrichment will be significantly delayed from the
onset of star formation, and alpha-enhancement will be associated
with short-lived periods of relatively recent star formation.  
Significant early iron production by ``prompt'' SNe Ia would modify
this conclusion, making it more difficult to explain alpha-enhancement
and supporting alternative explanations such as a modification of the
initial mass function.  

Future studies will certainly improve on our results; the Baryon
Oscillation Spectroscopic Survey \citep[BOSS,][]{SchlegelEtAl07}, part of
SDSS-III\footnote{http://www.sdss3.org/cosmology.php}, is perhaps the
most promising.  BOSS is taking spectra of all supernova hosts from
SDSS-SN, expanding our sample by a factor of $\sim$3.  While a control
sample is not naturally defined for these SN hosts, the untargeted
nature of SDSS-SN means that a control group could be built out of a
deep, volume-limited sample of galaxies like GAMA \citep{BaldryEtAl09}.
Such a sample could be used to significantly tighten our error
bars. 

Increasing the temporal resolution of our recovered DTD may be more
challenging. This will require better spectra, both of SN hosts and
control galaxies, and especially better stellar population synthesis
models.  The latter point is crucial: we had to combine all ages
younger than 420 Myr into a single bin because of biases introduced by
the modeling of dust and stellar spectra. Splitting this age bin will
allow us to better determine the timescale of ``prompt'' SNe Ia and
thus the initial masses of their progenitors.  

With better stellar population synthesis models, BOSS, GAMA, and other
surveys would hold tremendous promise.  They would allow us to
construct a sample several times as large as the one used here, and
with better recovered star formation histories.  With such a sample,
we could improve both the precision and temporal resolution of our
DTD by a factor of $\sim$2, placing strict constraints on
theoretical SN Ia progenitor models.  

\section{Acknowledgments}
Funding for the SDSS and SDSS-II has been provided by the Alfred
P. Sloan Foundation, the Participating Institutions, the National
Science Foundation, the U.S. Department of Energy, the National
Aeronautics and Space Administration, the Japanese Monbukagakusho,
the Max Planck Society, and the Higher Education Funding Council
for England. The SDSS Web Site is http://www.sdss.org/. 

The SDSS is managed by the Astrophysical Research Consortium for the
Participating Institutions. The Participating Institutions are the
American Museum of Natural History, Astrophysical Institute Potsdam,
University of Basel, University of Cambridge, Case Western Reserve
University, University of Chicago, Drexel University, Fermilab, the
Institute for Advanced Study, the Japan Participation Group, Johns
Hopkins University, the Joint Institute for Nuclear Astrophysics, the
Kavli Institute for Particle Astrophysics and Cosmology, the Korean
Scientist Group, the Chinese Academy of Sciences (LAMOST), Los Alamos
National Laboratory, the Max-Planck-Institute for Astronomy (MPIA),
the Max-Planck-Institute for Astrophysics (MPA), New Mexico State
University, Ohio State University, University of Pittsburgh,
University of Portsmouth, Princeton University, the United States
Naval Observatory, and the University of Washington.

We thank an anonymous referee for many corrections and helpful
suggestions.  We would also like to thank the entire SDSS
collaboration and especially the SDSS Supernova Survey team for their
work in building the exceptional dataset used in this paper.  

This material is based upon work supported under a National Science
Foundation Graduate Research Fellowship to TDB.  It was also supported in
part by  DOE grant DE-FG02-07ER41514.  RT thanks the UK Science and
Technology Facilities Council and the Leverhulme Trust for financial
support.  

\bibliographystyle{apj_eprint}
\bibliography{snIarefs}

\end{document}